\documentclass[10pt]{article}

\usepackage{amsmath,amssymb,amsfonts}
\usepackage{graphicx,subfigure}
\usepackage{cite}

\newcommand{\avg}[1]{\left\langle{#1}\right\rangle}
\newcommand{\avgq}[1]{\left\langle{#1}\right\rangle_{\boldsymbol{q}}}
\newcommand{\bfq}{\boldsymbol{q}}

\newcommand{\dudxs}{u'(x^*)}

\newcommand{\req}[1]{(\ref{#1})}
\newcommand{\ii}{\text{i}}

\renewcommand{\l}{\left}
\renewcommand{\r}{\right}

\newcommand{\beq}{\begin{equation}}
\newcommand{\eeq}{\end{equation}}
\newcommand{\beqa}{\begin{eqnarray}}
\newcommand{\eeqa}{\end{eqnarray}}

\begin{document}

\title{Typical properties of large random economies with linear
activities} 

\author{A. De Martino$^1$, M. Marsili$^2$ and I. P\'erez Castillo$^3$\\
$^1$INFM-SMC and Dipartimento di Fisica, Universit\`a di Roma ``La
Sapienza''\\P.le A. Moro 2, 00185 Roma (Italy)\\
$^2$The Abdus Salam International Centre for Theoretical Physics\\
Strada Costiera 14, 34014 Trieste (Italy)\\
$^3$Instituut voor Theoretische Fysica, Katholieke Universiteit Leuven\\
Celestijnenlaan 200D, 3001 Leuven (Belgium)}

\maketitle

\begin{abstract}
We study the competitive equilibrium of large random economies with
linear activities using methods of statistical mechanics. We focus on
economies with $C$ commodities, $N$ firms, each running a randomly
drawn linear technology, and one consumer. We derive, in the limit
$N,C\to\infty$ with $n=N/C$ fixed, a complete description of the
statistical properties of typical equilibria. We find two regimes, which in the limit of efficient technologies are separated by a phase transition, and argue that endogenous technological change drives the economy close to the critical point.
\end{abstract}


\section{Introduction}

The aggregation of microeconomic behavior into macroeconomic laws is a
difficult task because of the presence of heterogeneity both at the
level of individual characteristics and of interactions. The paradigm
of the representative agent, which essentially reduces the problem to
that of a single macro individual, has shown all its inadequacy
\cite{Kirman}, calling for alternative approaches. Computational
methods -- both in the spirit of agent based modeling or implementing
general equilibrium theory -- represent a viable substitute, rapidly
growing in popularity. However these techniques provide punctual
results which are difficult to generalize. While they are very useful
in deriving specific results for a specific economy, they do not
lead to a broad understanding.

At the other extreme, the methods of mathematical economics aim at
general results -- such as existence, uniqueness, efficiency -- which
hold for broad classes of situations. Pinning down the typical
macroeconomic behavior beyond these general results is however very
hard, especially when agents are heterogeneous (e.g. in their
endowments, technologies, budgets, utility functions, \ldots) and are
interconnected via a complex network of interactions.

Understanding the complex macro-behavior of a system does not
necessarily require a detailed description of it in all its
complications. Indeed many laws which govern macro-behavior have a
statistical origin. E.P. Wigner \cite{Wigner} first had the intuition
that in such cases, the collective behavior of a large system with $N$
degrees of freedom -- heavy atoms in his case -- is well approximated
by that of a system with random interactions in the limit
$N\to\infty$. Indeed, if the relevant properties obey laws of large
numbers, then they will be substantially independent of the specific
realization of the interactions when $N$ is large.  

The statistical properties of random systems have been a central
research issue in statistical mechanics for the past two decades, and
extremely powerful analytical tools to calculate them have been
developed. These techniques have already found a wide range of
applications outside physics: among others, in combinatorial
optimization problems and computer science, in the theory of neural
networks, in information theory, and in agent based models
\cite{matching,ksat,hkp,cmz,bmrz}. 

It has been realized several times by different authors \cite{Foley,Durlauf} that tools developed in statistical mechanics can be useful in economic theory. While the idea of studying large random economies as a proxy for a complex economy with heterogeneous agents may not be entirely new (see e.g. \cite{Follmer}), modern tools of statistical mechanics of disordered systems have not yet been exploited. Here we apply these tools to the study of the typical properties of large random production economies. The model we shall consider, outlined in detail in the next section, is based on a $C$ dimensional commodity space and
has $N$ firms with linear technologies, as in \cite{Lancaster}.
Feasible technologies are assumed to be drawn at random from some
probability distribution. Firms chose technologies from the set of
feasible ones and fix the scale of operations so as to maximize their
respective profits. The total supply is matched to the demand of a
single consumer with initial endowments drawn at random from a given
distribution. Equilibrium prices, operation scales and consumption
levels are determined by imposing that all markets clear. Equilibrium
quantities are random, as they depend on the draw of technologies and
o initial endowments.  A complete statistical characterization of an
ensemble of equilibria of large random economies is obtained in the
limits $N\to\infty$ and $C\to\infty$ . The laws we derive are of a
statistical nature, i.e. laws that a typical realization of an economy
from the ensemble will satisfy almost surely, i.e. with a probability
close to one when $N$ is large.

We show in particular that this approach: i) identifies the
relevant macroscopic variables, the so-called {\em order parameters},
describing the behavior of the system in the limit $N\to\infty$;
ii) allows the calculation of the values of the order parameters from
the solution of a ``representative'' firm problem, which embodies all
the complexity of the full heterogeneous model; iii) enables one to
derive distributions of consumption levels and of scales of activities
at equilibrium. We will prove that for a broad class of
choices the properties of the competitive equilibria change
qualitatively at a critical value $n_c=2$ of the ratio $n=N/C$. This change becomes a sharp phase transition in the limiting
case of efficient technologies.
Loosely speaking, the economy expands rapidly when $n$ increases for $n<2$ whereas, when $n>n_c$ the economy is in a mature phase where the technology space is to a large extent saturated. Even though our picture is static, we shall claim, in the final section, that in a dynamic setting technological innovation (i.e. changes in $N$ and/or $C$) driven by total output growth are likely to drive the economy close to the critical value $N/C=2$.

After discussing the model in the next Section we present, in Section
3, the main results. In order not to obscure the emergent picture, a
detailed account of the approach and of the calculation is given in
the appendix. More specific and quantitative results will be discussed
in Section 4 and in Section 5 we argue that economies self-organize close to the critical point $n\approx 2$. 
We close by summarizing our results and discussing some perspectives in the final section.

We made an effort to keep the discussion and the mathematical
complexity at the simplest level, even at the price of introducing
restrictive or unrealistic assumptions. The present approach can
however be easily generalized to more realistic (and more complicated)
models.

\section{The model} 
 
We consider an economy with $C$ commodities, $N$ firms endowed with 
random technologies, and one representative consumer. Firms strive to 
maximize their respective profits, while the consumer aims at 
maximizing his utility. The two problems are 
interconnected by the market clearing condition. In detail, we 
consider the following setup. 
 
The company $i$ ($i=1,\ldots,N$) is characterized by a technology (or
activity) with constant returns to scale that, when run at scale $s_i=1$
produces $q_i^c>0$ or consumes $q_i^c<0$ units of commodity $c$
($c=1,\ldots,C$). If the technology $\boldsymbol{q}_i
=\{q_i^c\}_{c=1}^C$ is operated at a scale $s_i>0$, then firm $i$
produces or consumes $s_i q^c_i$ units of commodity $c$. Technologies
cannot be reversed, i.e. $s_i\geq 0$. Following \cite{Lancaster}, we
do not restrict our attention to Leontiev input-output models: each
activity can have several outputs (joint production) and there are no primary production
factors (i.e. $q_i^c>0$ is possible for all $c$). As in
\cite{Lancaster}, it will be important to impose that it is impossible
to produce a positive amount of some commodity without consuming a
positive amount of some other commodity. A sufficient condition to
ensure this is that
\begin{equation} 
\forall i:\qquad\sum_{c=1}^C q_i^c=-\epsilon .
\label{constraint} 
\end{equation} 
Here $\epsilon$ is the difference between the quantities of inputs
and outputs, which measures the inefficiency of the transformation
process of technology $i$.
 
The profit of firm $i$ is $\pi_i=
s_i(\boldsymbol{p}\cdot\boldsymbol{q}_i) \equiv s_i \sum_{c=1}^C
q^c_{i} p^c$, where $\boldsymbol{p}=\{p^c\}_{c=1}^C$ is the price
vector, which we assume to be non-negative. Each firm fixes $s_i$ by
solving the problem
\begin{equation} 
\max_{s_i\geq 0}~\pi_i 
\label{maxf} 
\end{equation} 
at fixed prices.  
 
The representative consumer, whose utility function we denote by
$U(\cdot)$ and whose initial endowment we denote by
$\boldsymbol{x}_0$, chooses his consumption
$\boldsymbol{x}=\{x^c\}_{c=1}^C$ by solving
\begin{equation}\label{maxc} 
\max_{\boldsymbol{x}\in {\cal B}}~U(\boldsymbol{x}), \qquad {\cal 
B}=\{x^c\ge 0:\boldsymbol{p}\cdot\boldsymbol{x}\le 
\boldsymbol{p}\cdot\boldsymbol{x}_0\} 
\end{equation} 
at fixed prices. ${\cal B}$ represents the set of consumption plans that 
satisfy the consumer's budget constraint. 
 
At equilibrium, the total supply of each commodity is required to 
match the demand from the representative consumer (market clearing), 
i.e. 
\begin{equation}\label{mc} 
\forall c:\qquad x^c=x_0^c+\sum_{i=1}^N s_i q_i^c  
\end{equation} 
The simultaneous solution of the maximization problems (\ref{maxf}) 
and (\ref{maxc}) subject to \req{mc} constitutes the competitive 
equilibrium we will study in this paper. 
 
Before specifying further our model, it is worth to make a couple of
remarks. First, multiplying both sides of (\ref{mc}) by $p^c$ and
summing over $c$ one finds that, in equilibrium,
\begin{equation}\label{Walras}
\boldsymbol{p}\cdot(\boldsymbol{x}-\boldsymbol{x}_0)=\sum_{i=1}^N s_i 
(\boldsymbol{p}\cdot\boldsymbol{q}_i)=\sum_{i=1}^N\pi_i=0 
\end{equation} 
The last equality comes from the fact that 
$\boldsymbol{p}\cdot(\boldsymbol{x}-\boldsymbol{x}_0)\le 0$ because of 
the budget constraint and $\pi_i\ge 0$ because firms can always 
achieve $\pi_i=0$ by not producing. So, on one side one recovers 
Walras' Law $\boldsymbol{p}\cdot(\boldsymbol{x}-\boldsymbol{x}_0)= 0$, 
while on the other we find that $\pi_i=0$ for all firms. 
 
Notice also that, combining \req{constraint} with the market clearing 
condition we find that in equilibrium 
\begin{equation} 
\sum_{c=1}^C (x^c-x_0^c)=-\epsilon\sum_{i=1}^N s_i 
\label{selfcons} 
\end{equation} 
This equation means that total equilibrium consumption will be lower 
than the initial one. The model thus focuses on the ability of the 
productive sector to provide scarce goods (with small $x_0^c$) using 
as inputs abundant commodities (with large $x_0^c$) so as to increase 
welfare. 
 
We assume that technologies $q_i^c$ are given by
\begin{equation}\label{pdiq}
  q_i^c=r_i^c-\frac{\epsilon}{C}-\frac{1}{C}\sum_{c=1}^C r_i^c
\end{equation}
where $r_i^c$ are independent Gaussian random variables, with zero mean and
variance $\Delta/C$ and the last two terms enforce the constraint
\eqref{constraint}. Appendix \ref{appconti} shows that the assumption
on the distribution of $q_i^c$ can be relaxed considerably for our
purposes\footnote{In fact, any distribution which satisfies
\eqref{constraint} and has a characteristic function
$\log\avgq{e^{ikq_i^c}}=\psi(k/\sqrt{C})$ with $\psi(x)=-\Delta
x^2/2+O(x^3)$ would leave our results unchanged.}.

In what follows, we shall use the notation
$\avg{\ldots}_{u,v,\ldots,z}$ for expected values over the
distributions of the variables $u,v,\ldots,z$, but we shall omit the
subscript when no confusion is possible.

Commodities are {\em a priori} equivalent. The initial endowments
$x_0^c$ are drawn at random from a distribution $\rho(\cdot)$,
independently for each $c$. Furthermore we shall also suppose that
\begin{equation}  
U(\boldsymbol{x})=\sum_{c=1}^C u(x^c) 
\label{ansatzu} 
\end{equation} 
where $u(\cdot)$ is postulated to be increasing ($u'(x)>0$) and convex
($u''(x)<0$). These assumptions, which simplify our analysis
considerably, appear to be extremely restrictive. They appear less unrealistic considering, as in \cite{Lancaster66}, that $\boldsymbol{x}$ may measure desirable characteristics or properties of commodities rather than quantities thereof. In this light, the departure from Leontiev
technologies with a single output becomes natural.

It is also useful to introduce a measure of economic activity similar to the Gross Domestic Product (GDP). The total market value of all goods produced 
is the sum of $(x^c-x_0^c)p^c$ for all $c$ with $x^c>x_0^c$. Because of Walras' law \req{Walras}, this is equal to $\frac{1}{2}\sum_{c}|x^c-x_0^c|p^c$. Normalizing prices to the average price level, we obtain
\begin{equation}\label{gdp}
  {\rm GDP}=C\frac{\sum_{c=1}^C|x^c-x_0^c|p^c}{2\sum_{c=1}^C p^c}.
\end{equation}

The key parameters of the model are thus $N,~C,~\epsilon,~\Delta$, the
distribution $\rho(\cdot)$ of the initial endowments, and the utility
function $u(\cdot)$. We shall focus on the non trivial limit
$N\to\infty$ defined as
\begin{equation}\label{lim} 
\lim_{N\to\infty}^{(n)}\equiv \lim_{\substack{N\to\infty\\n=N/C}} 
\end{equation} 
where $n=N/C$ is held fixed as $N\to\infty$. 

A simple geometric argument, for $\epsilon=0$, suggests that $n=2$ will
play an important role. Let us write the initial endowments as
$x_0^c=\bar x_0+\delta x_0^c$, separating a constant part
($\bar x_0$) from a fluctuating part ($\delta x_0^c$) such that $\sum_c\delta x^c=0$. With
$\epsilon=0$, Eq. \eqref{selfcons} implies that the component of
consumption along the constant vector remains constant.  All the
transformations take place in the space orthogonal to the constant
vector: $\bfq_i\cdot \boldsymbol{x}_0 = \bfq_i\cdot \boldsymbol{\delta
x}_0$. In other words, those technologies with $\bfq_i\cdot
\boldsymbol{\delta x}_0<0$ which reduce the initial spread of
endowments $\boldsymbol{\delta x}_0$ lead to a increase in wealth and
hence will be run at a positive scale. Those with a positive component
along $\boldsymbol{\delta x}_0$ will have $s_i=0$. Given that the
probability to generate randomly a vector in the half-space $\{\bfq
:~\bfq\cdot \boldsymbol{\delta x}_0<0\}$ is $1/2$, when $N$ is large
we expect $N/2$ active firms. Still the number of possible active
firms is bounded above by $C$, hence when $n=N/C=2$ the space of
technologies becomes complete and $x^c=\bar x_0$~$\forall c$. There is
no possibility to increase welfare further. We shall see that $n=2$ separates two distinct regimes of equilibria even with $\epsilon>0$.

It is easy to see that the problem of finding equilibrium prices, 
production scales and consumption levels of the economy in the 
above setting is reduced to the following: 
\begin{equation}\label{pro} 
\max_{\{s_i\ge 0\}}U\left(\boldsymbol{x}_0+\sum_{i=1}^Ns_i 
\boldsymbol{q}_i\right)  
\end{equation} 
Given the solution $\{s_i^*\}$ to this problem, the equilibrium 
consumption levels are given by the market clearing condition 
(\ref{mc}) and the (relative) prices are derived from marginal 
utilities as\footnote{Utility maximization under the budget constraint, Eq. \eqref{maxc}, yields $\partial U/\partial x^c=\lambda p^c$ where $\lambda$ is the Lagrange multiplier imposing the budget constraint. We can take $\lambda=1$ exploiting the invariance $p^c\to a p^c$ for any $a>0$, thus
fixing the level of absolute prices.}
\begin{equation}\label{pric} 
p^c=\left. \frac{\partial U}{\partial 
x^c}\right|_{\boldsymbol{x}^*}=u'({x^*}^c) 
\end{equation} 

Eq. \req{pro} is a typical problem in statistical mechanics. The
general approach to this type of issues is discussed in Appendix
\ref{method}. Equilibrium quantities are random variables because of
the randomness in the technologies $\boldsymbol{q}_i$ and in the
initial endowments $\boldsymbol{x}_0$. Still there are statistical
properties of the equilibrium which hold almost surely in the limit
\req{lim}. These will be the subject of our interest. In Section
\ref{solution}, we present the general solution, while in Section
\ref{discussion} we shall specialize to specific examples. The
reader interested in technical details is referred to the appendices
for a detailed account of the method and of the explicit calculation.
 
\section{The solution and its generic properties} 
\label{solution} 
 
As shown in appendices \ref{method} and \ref{appconti}, the solution of the equilibrium problem \req{pro} in the limit $N\to\infty$ with $n=N/C$ fixed is given by
\begin{equation} 
\lim_{N\to\infty}^{(n)} 
\frac{1}{N}\avgq{\max_{\boldsymbol{s}}~U\left(\boldsymbol{x}_0+\sum_i 
s_i\boldsymbol{q}\right)}= h(\Omega^*,\kappa^*,p^*,\sigma^*,\chi^*,\hat\chi^*) 
\end{equation} 
where
\begin{multline}\label{h} 
h(\Omega,\kappa,p,\sigma,\chi,\hat\chi)= \left\langle \max_{s\ge 
0}\left[(t\sigma-\epsilon p)s-\frac{1}{2}\hat\chi 
s^2\right]\right\rangle_t +\frac{1}{2}\Omega\hat\chi+\frac{1}{n}\kappa 
p-\frac{1}{2n\Delta}\chi\sigma^2- \frac{1}{2 n}\chi p^2+\\+ 
\frac{1}{n} \avg{\max_{x\ge 
0}\left[u(x)-\frac{1}{2\chi}\left(x-x_0+\kappa+\sqrt{n\Delta \Omega} 
t\right)^2\right]}_{t,x_0} 
\end{multline} 
and $\Omega^*,\ldots,\hat\chi^*$ are the saddle point values of the parameters, i.e. those which solves the system of equations 
$\frac{\partial h}{\partial \Omega}=0,\ldots, \frac{\partial h}{\partial \hat\chi}=0$. The variables
$\Omega,\kappa,p,\sigma,\chi,\hat\chi$ are called {\em order
parameters} in statistical physics\footnote{In order to keep notation simple, we shall generally omit in what follows the asterisk on order parameters which denotes saddle point values.}. They emerge from the analytic approach (see appendices \ref{method} and \ref{appconti}) as the key macroscopic variables which describe the collective behavior of the equilibria.

In Eq. \eqref{h} $t$ is a Gaussian r.v., with zero mean and unit
variance and as usual $\avg{\ldots}_t$, $\avg{\ldots}_{t,x_0}$ stand
for expectation values on $t$ and on $t$ and $x_0$, respectively. The
precise derivation of this result is described in Appendix
\ref{appconti}.
 
The structure of $h$ is reminiscent of the original problem. The first
term on the r.h.s. can indeed be regarded as the profit maximization
of a ``representative'' firm. The variable $s$ is indeed one of the
variables $s_i$ which appear in the original problem \eqref{pro}. The
solution of the maximization problem in the first term of
Eq. \eqref{h} is given by
\begin{equation}\label{sstar} 
s^*(t)= 
\begin{cases} 
(t\sigma-\epsilon p)/{\hat\chi}&\text{if $t\ge \epsilon p/\sigma$}\\ 
0&\text{if $t< \epsilon p/\sigma$} 
\end{cases} 
\end{equation} 
Since $t$ is a random variable, $s^*$ is also a random variable and
its probability density can be derived from that of $t$. The result is
\begin{equation} 
Q(s)=(1-\phi)\delta(s)+ 
\frac{\hat\chi}{\sqrt{2\pi}\sigma}\Theta(s)\exp\l[-\frac{(\hat\chi 
s+\epsilon p)^2}{2\sigma^2}\r],\qquad \phi=\frac{1}{2}{\rm 
erfc}\left(\frac{\epsilon p}{\sqrt{2}\sigma}\right)
\label{pdis} 
\end{equation} 
where $\Theta(s)=0$ for $s\le 0$ and $\Theta (s)=1$ for $s>0$. The
variable $s$ is the scale of production of a (representative) firm,
hence \req{pdis} yields the distribution of $s_i$ in the economy and
$\phi$ is the fraction of technologies that are active (i.e. such that
$s_i>0$).
 
Likewise, the last term on the r.h.s. of Eq. \req{h} is related to
utility maximization with respect to a ``representative''
commodity. The variable $x$ is indeed one of the variables $x^c$ which
appear in the original problem. The solution of this problem is given
by
\begin{equation} 
x^*(t,x_0):\qquad \chi 
\dudxs 
=x^*-x_0+\kappa+\sqrt{n\Delta \Omega}t 
\label{xstar} 
\end{equation} 
which is always positive provided $u'(x)\to\infty$ for $x\to 0$. The
probability density of $x^c$ in the economy can be derived from that
of $t$ and $x_0$ in the same way as above for the scale $s_i$ of
production. The conditional probability of $x^c$ given $x_0^c$ is
computed in Appendix \ref{apppdf}. The result is
\begin{equation} 
\label{pdix} 
P(x|x_0)= \frac{1-\chi u''(x)}{\sqrt{2\pi n\Delta \Omega}} 
\exp\left\{-\frac{\left(x-x_0-\chi u'(x)+\kappa\right)^2} {2n\Delta 
\Omega}\right] 
\end{equation} 
Hence the variable $x-\chi u'(x)$ has a Gaussian distribution with
mean $x_0-\kappa$ and variance $n\Delta \Omega$.
 
The two ``representative'' problems are coupled in a nontrivial way
through the other terms in \req{h}.
 
The structure of the solution becomes more clear if we analyze the set
of saddle point equations $\frac{\partial h}{\partial \Omega}=0,\ldots, \frac{\partial h}{\partial \hat\chi}=0$, with
$\boldsymbol{\theta}=(\Omega,\kappa,p,\sigma,\chi,\hat\chi)$. After
some algebra (see appendix \ref{appconti}), these can be cast in the following form:
\begin{eqnarray} 
p&=&\left\langle\dudxs\right\rangle_{t,x_0} 
\label{p}\\ 
\hat\chi&=&\sqrt{\frac{\Delta}{n\Omega}} 
\left\langle\dudxs t\right\rangle_{t,x_0} 
\label{chihat}\\ 
\sigma &=& \sqrt{\Delta\left[\left\langle\left(\dudxs\right)^2  
\right\rangle_{t,x_0}-\left\langle\dudxs\right\rangle_{t,x_0}^2\right]} 
\label{sigma}\\ 
\Omega&=&\avg{(s^*)^2}_t 
\label{Omega}\\ 
\chi &=& \frac{n\Delta}{\sigma}\avg{s^* t}_t 
\label{chi}\\ 
\kappa &=& p\chi+n\epsilon\avg{s^*}_t 
\label{k} 
\end{eqnarray} 
The first of these equations relates the parameter $p$ to the average
(relative) price because of \req{pric}, while the third one implies
that $\sigma$ is a measure of price fluctuations\footnote{It is 
possible to derive an explicit analytic form of (\ref{Omega}),
(\ref{chi}) and \req{k} in terms of error functions. The present
formulas are however more suited for the discussion which follows.}.
 
Using these relations (see appendix \ref{appgeneric} for details) one
finds that at the saddle point
\begin{equation} 
h(\Omega^*,\kappa^*,p^*,\sigma^*,\chi^*,\hat\chi^*)=\frac{1}{n}\avg{u(x^*)}_{t,x_0} 
\label{hsp} 
\end{equation} 
This is indeed what we expect looking at the original problem
\req{pro}. Furthermore, taking the expected value of \req{xstar} and
combining it with (\ref{p}) and (\ref{k}) yields
$\chi\avg{\dudxs}_{t,x_0}=\chi
p=\avg{x^*}_{t,x_0}-\avg{x_0}_{x_0}+\kappa=
\avg{x^*}_{t,x_0}-\avg{x_0}_{x_0}+p\chi +n\epsilon\avg{s^*}_t$. Then
\beq \avg{x^*}_{t,x_0}=\avg{x_0}_{x_0}-n\epsilon\avg{s^*}_t
\label{avgx} 
\eeq which is exactly Eq. \req{selfcons}. Finally, it is possible to show
(see Appendix \ref{appgeneric}) that Eq.s (\ref{p}--\ref{k}) also
``contain'' Walras' law in the form
\begin{equation} 
\avg{\dudxs(x^*-x_0)}_{t,x_0}=0. 
\label{walras} 
\end{equation} 
The dependence on $\Delta$ of the solution can be clarified by a
rescaling argument: changing variables to $p'=p$,
$\hat\chi'=\hat\chi/\Delta$, $\sigma'=\sigma/\sqrt{\Delta}$,
$\Omega'=\Delta \Omega$, $\chi'=\chi$ and $\kappa'=\kappa$ one finds
that the solution only depends on the parameter
$\epsilon'=\epsilon/\Delta$. Hence the behavior of the solution with respect to $\Delta$ is easily related to the dependence on $\epsilon$ with $\Delta=1$.
Notice that, a dependence on $\Delta$ remains after the change of variables in the distribution of $s_i$, Eq. \eqref{pdis}. This means that production scales satisfy the scaling relation 
\beq 
s_i(\Delta)=s_i(1)/\sqrt{\Delta}
\label{scaling}
\eeq 

\noindent
The behavior of the solution when the spread of the initial endowments
$\avg{\delta x_0^2}\equiv\avg{(x_0-\avg{x_0})^2}$ is very
small can be computed with asymptotic expansion methods. The key
observation in the expansion (see appendix \ref{appexpdx})~ is that
$x^*$ also has very small fluctuations. This, in turn, implies that
prices also have very small fluctuations, indeed $\sigma\cong
|u''(\avg{x_0})|\sqrt{\avg{\delta x_0^2}}$. The scales of production
also vanish when $\avg{\delta x_0^2}\to 0$, but with a singular
exponential behavior:
\begin{equation}
\Omega\propto \avg{\delta x_0^2}^{3/2}e^{-A/\avg{\delta
x_0^2}},~~~~\avg{\delta x_0^2}\ll 1
\end{equation}
for some constant $A$. Hence we find that no economic activity takes
place ($\phi\to 0$, $\Omega\to 0,\avg{s^*}_t\to 0$) in the limit of
uniform endowments (see Appendix \ref{appexpdx} for technical details). This is what one should expect from the beginning: 
when the consumer is endowed with the same amount of equally valued commodities, there is no transformation (with $\epsilon\ge 0$) which can increase welfare.

A further interesting limit, for which we can derive generic results,
is that of vanishing $\epsilon$. Setting $\epsilon=0$ one finds in a straightforward way that
$\Omega=\sigma^2/(2\hat\chi^2)$, $\chi=n\Delta/(2\hat\chi)$ and
$k=p\chi$. Eq. \eqref{pdiq} yields $\phi=1/2$ which means that half of
the firms are active, in agreement with the geometric argument of the
previous section for $n<2$. When $n\to 2^-$ the equations develop a
singularity: Indeed $\hat\chi\propto(2-n)$, $\sigma\propto \sqrt{2-n}$
vanish whereas the average scale of production diverges $\avg{s^*}\propto
1/\sqrt{2-n}$. A detailed account is given in Appendix
\ref{appexpeps}. The case $n>2$ is more subtle as it requires a
careful asymptotic study of the limit $\epsilon\to 0$ where again
realizing that $x^*$ has small fluctuations of order $\epsilon$ is
crucial. The bottom line is that (see appendix \ref{appexpeps} for
details) price fluctuations vanish linearly with $\epsilon$,
i.e. $\sigma\propto\epsilon$ but also $\hat\chi\propto \epsilon$ so
the factors $\epsilon p/\hat\chi$ and $\sigma/\hat\chi$ in
Eq. \eqref{sstar} are finite. Hence scales of production remain
finite, as $\epsilon\to 0$ and they diverge when $n\to 2^+$ as $\avg{s^*}\propto 1/\sqrt{n-2}$. The
fraction of active firms turns out to be $\phi=1/n$, which means that
there are exactly $C$ firms operating.

Eq.s (\ref{hsp}), (\ref{avgx}) and \req{walras} show that the saddle
point equations, which represent the simplest mathematical description
of the random economy in its full complexity, manage to capture in a
compact, though somewhat intricate, way the basic properties of the
economy. This is a useful consistency check. The best way to unravel
the resulting behavior beyond these generic laws is however to
specialize to particular cases.
 
\section{The solution: typical cases} 
\label{discussion} 
 
In this section we display the behavior of the solution outlined in
the previous section for some particular choices of the functions
$u(x)$ and $\rho(x_0)$. In spite of their apparent complexity, Eq.s
(\ref{p}--\ref{k}) can be solved numerically to any desired degree of
accuracy. Using the scaling argument above, we can safely restrict
ourselves to study the dependence on $\epsilon$ setting $\Delta=1$,
without any loss of generality.

\begin{figure} 
\begin{center} 
\includegraphics[width=8cm]{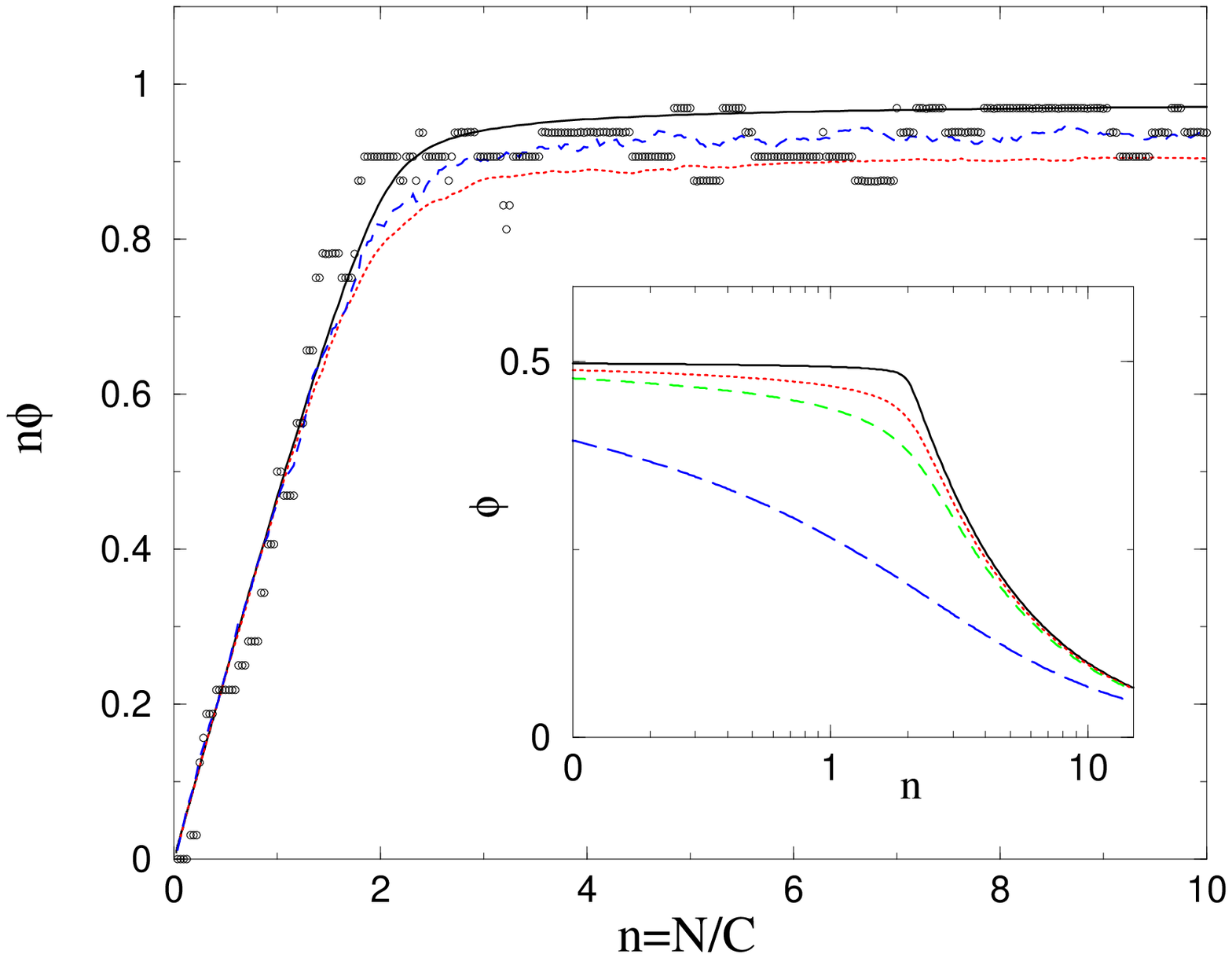} 
\includegraphics[width=8cm]{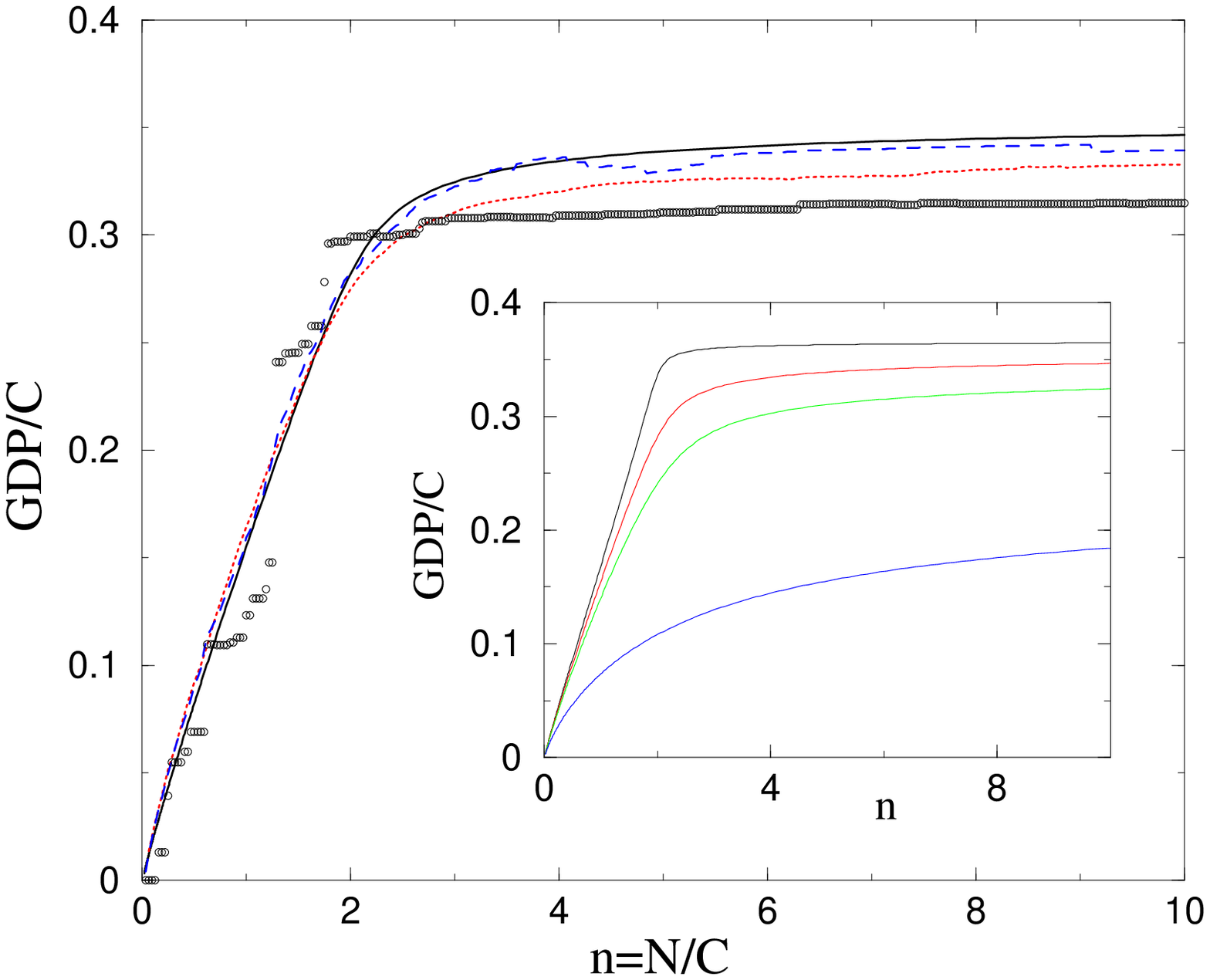} 
\caption{\label{fignum}Comparison between the analytic solution for 
$N\to\infty$ (full line) and equilibria of random economies computed 
numerically for $C=16$ and $32$. The parameters are $\epsilon=0.05$ and 
$\Delta=1$ whereas initial endowments are drawn from an exponential 
distribution. Dots refer to a single realization with $C=32$ whereas the dotted (dashed) line is the average over $100$ realizations for $C=16$ ($32$). Left: $n\phi$, which is 
the number of active firms ($s_i>0$) divided by $C$, versus 
$n$. Right: GDP versus $n$. Insets in these figures show the behavior of $\phi$ and of GDP for $\epsilon=0.01,~0.05,~0.1$ and $0.5$ from top to bottom.} 
\end{center} 
\end{figure}

We shall henceforth set \beq u(x)=\log x.  \eeq We start our 
discussion from the case \beq \rho(x_0)=e^{-x_0},\qquad x_0\ge 0. 
\eeq Fig.~\ref{fignum} compares the numerical solution with computer 
experiments.  We generate many realizations of the random economy and 
compute numerically the equilibria for each of them. The analytical 
results we obtain in the limit $C\to\infty$ turn out to give a quite 
accurate description of the behavior of relatively small 
systems\footnote{We resorted to a simple iterative scheme to converge 
to the equilibria. This fails to converge properly for $C$ or $N$ too 
large or for $\epsilon\ll 1$.} (i.e. $C= 16$) even for a single realization. 
Fig. \ref{fignum} shows that there are essentially two different 
regimes. For $n<n_c= 2$ roughly half of the firms 
are active, whereas for $n\gg n_c$ the number of active firms 
saturates to $C$. The GDP also shows a similar behavior. It increases with $n$ and saturates for $n>2$. 
 
\begin{figure} 
\begin{center} 
\includegraphics[width=8cm]{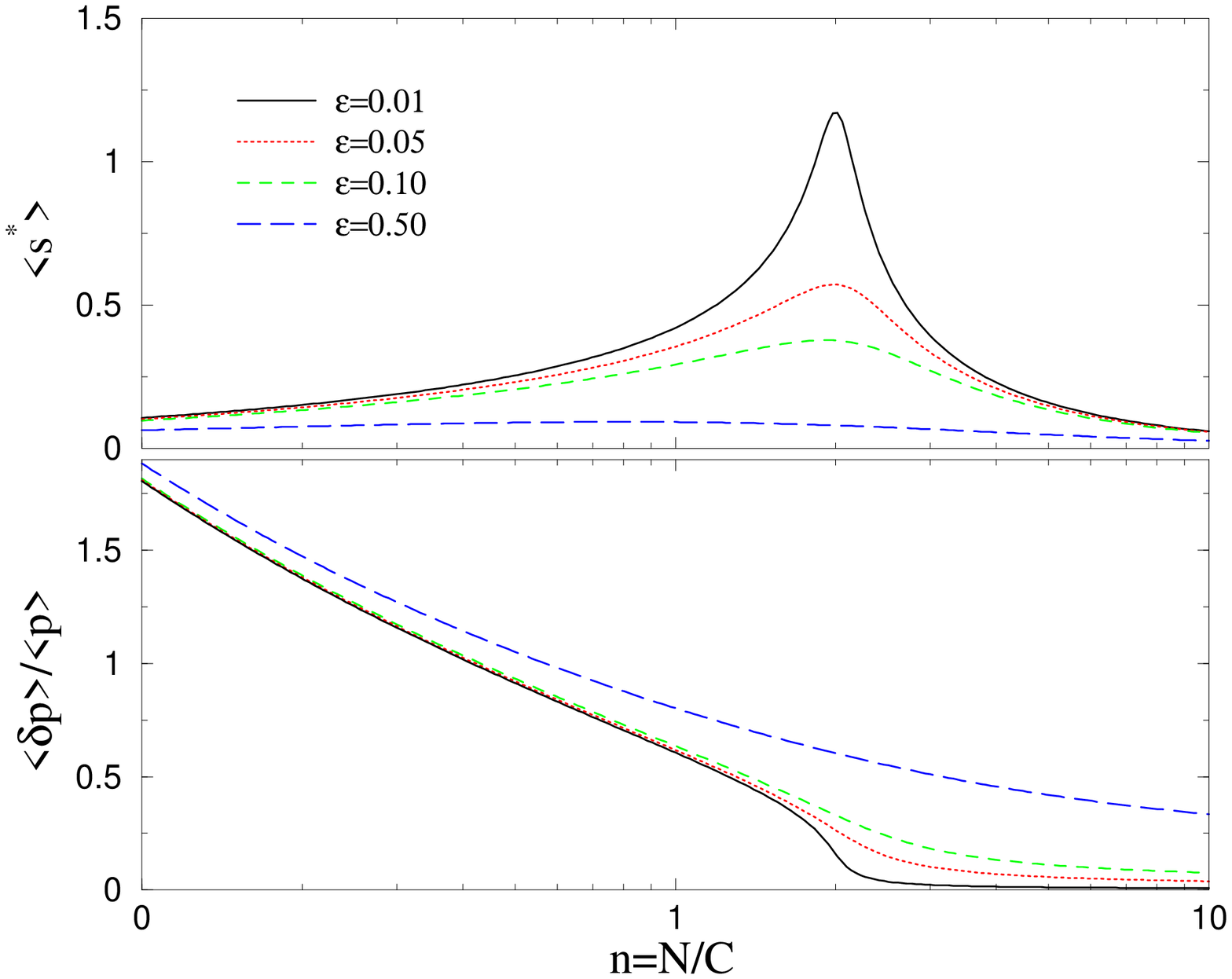} 
\includegraphics[width=8.2cm]{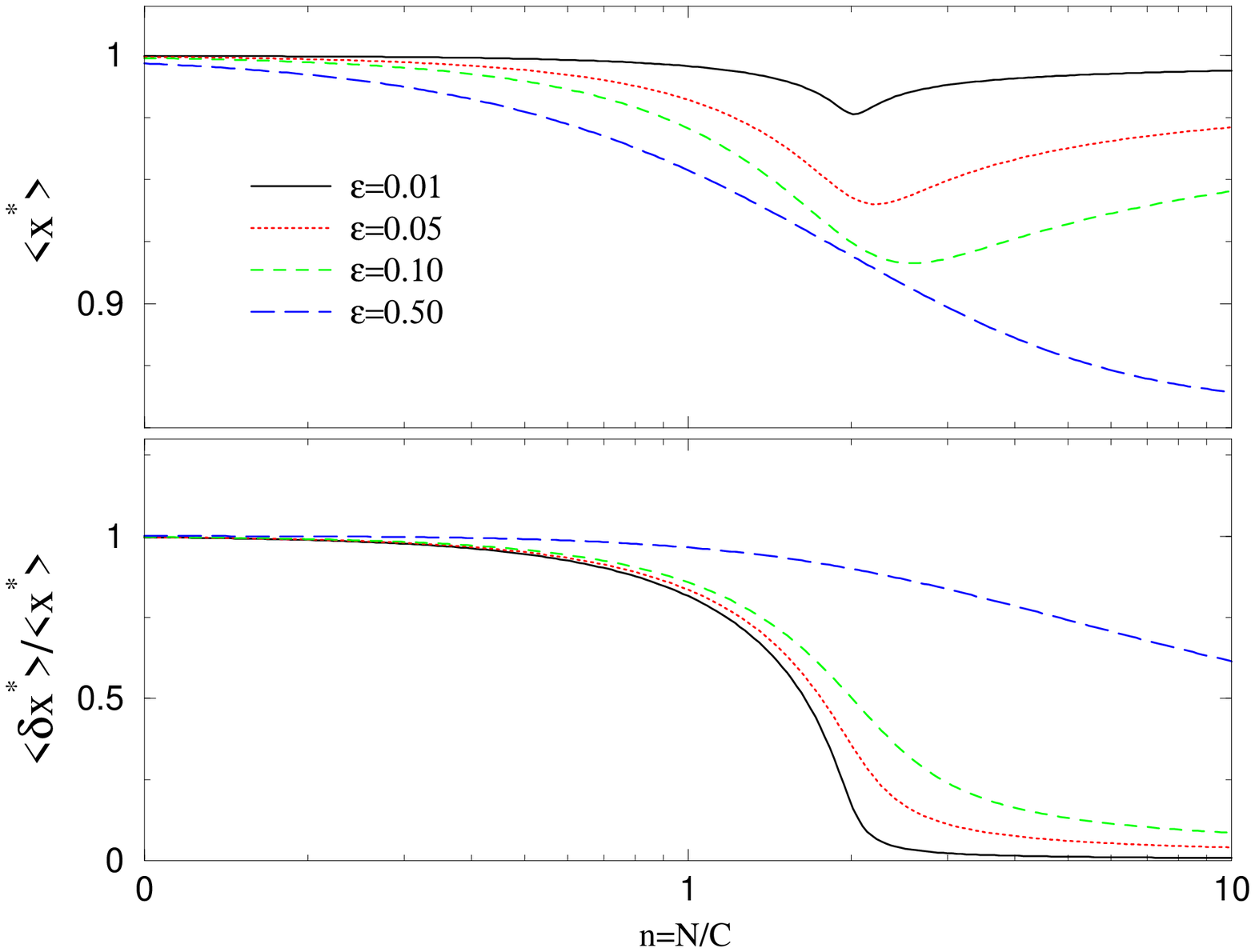} 
\caption{\label{sdpxdx}Behavior of equilibrium quantities as a 
function of $n$ for $\epsilon=0.5,~0.1,~0.05$ and $0.01$. In all cases 
$\Delta=1$ and $\rho(x)=e^{-x}$. Top left: $\avg{s^*}$. Bottom left: 
relative price fluctuations. Here we identify prices with marginal 
utility $p=u'(x^*)$ and $\delta p =u'(x^*)-\avg{u'(x^*)}$. Top right: 
average consumption $\avg{x^*}$. Bottom right: relative fluctuations 
of consumption $\delta x=x^*-\avg{x^*}$.} 
\end{center} 
\end{figure}

The transition between the two regimes becomes sharper when $\epsilon$
decreases and it gives rise to a singularity in the limit $\epsilon
\to 0$, as we have seen in the previous section. This is clearly
visible in Fig. \ref{sdpxdx}, where we plot the behavior of various
quantities as a function of $n$ for different values of $\epsilon$.
 
For $n<n_c$ the average scale of production $\avg{s^*}$ increases with
$n$. This means that, in this region, existing firms benefit from the
entry of a new technology (i.e. if $N\to N+1$, see later). This positive complementarity arises because the new firm increases the availability of inputs to other firms.

For $n>n_c$, instead $\avg{s^*}$ decreases with $n$, the introduction of a new technology typically causes a reduction in the scale of
activity of the already existing firms. When $\epsilon\to 0$ the curves develop a singularity $\avg{s^*}\sim 1/\sqrt{|n-2|}$ at $n_c=2$, as discussed in the previous section.
 
\begin{figure} 
\begin{center} 
\includegraphics[width=8cm]{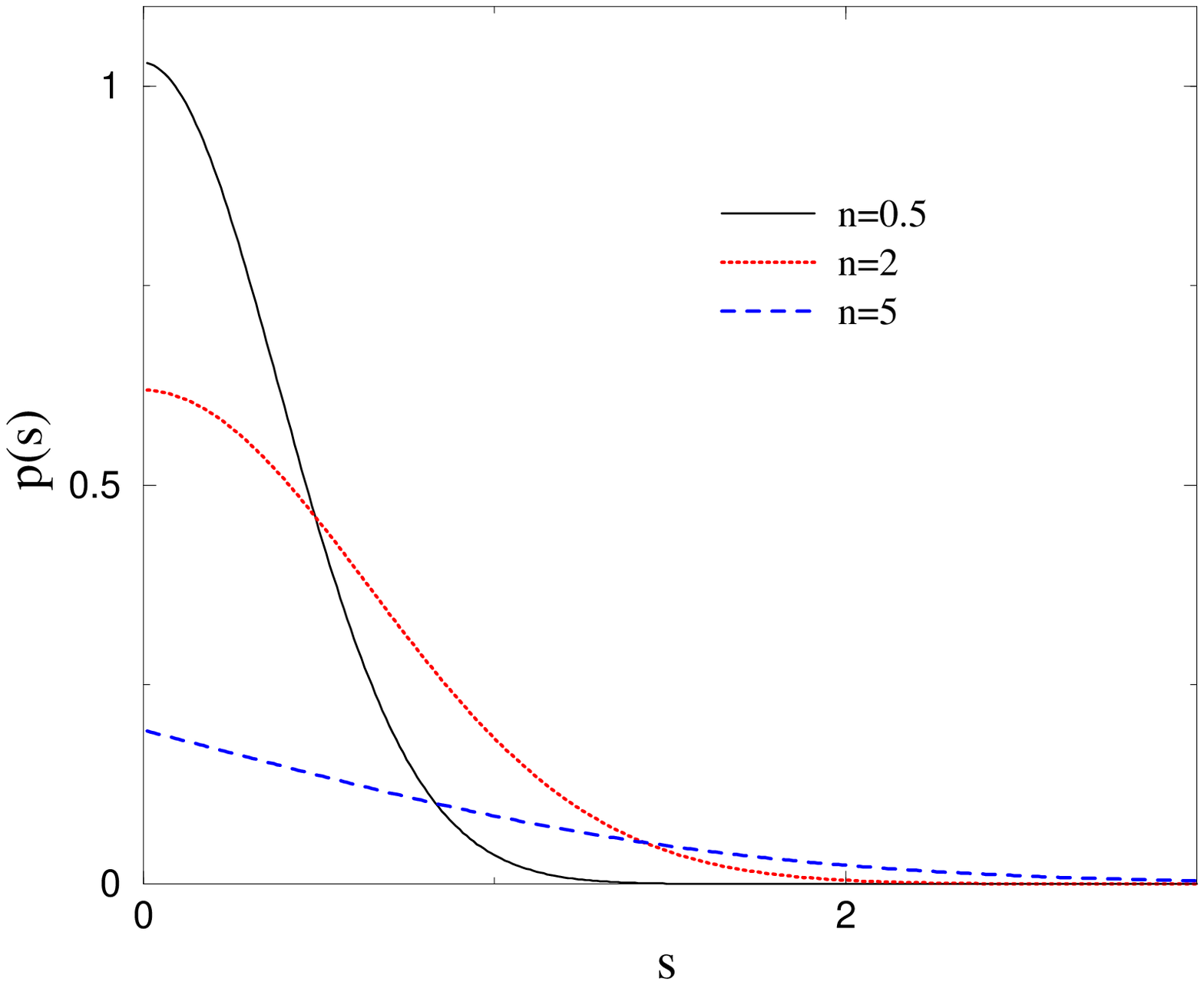} 
\includegraphics[width=8.2cm]{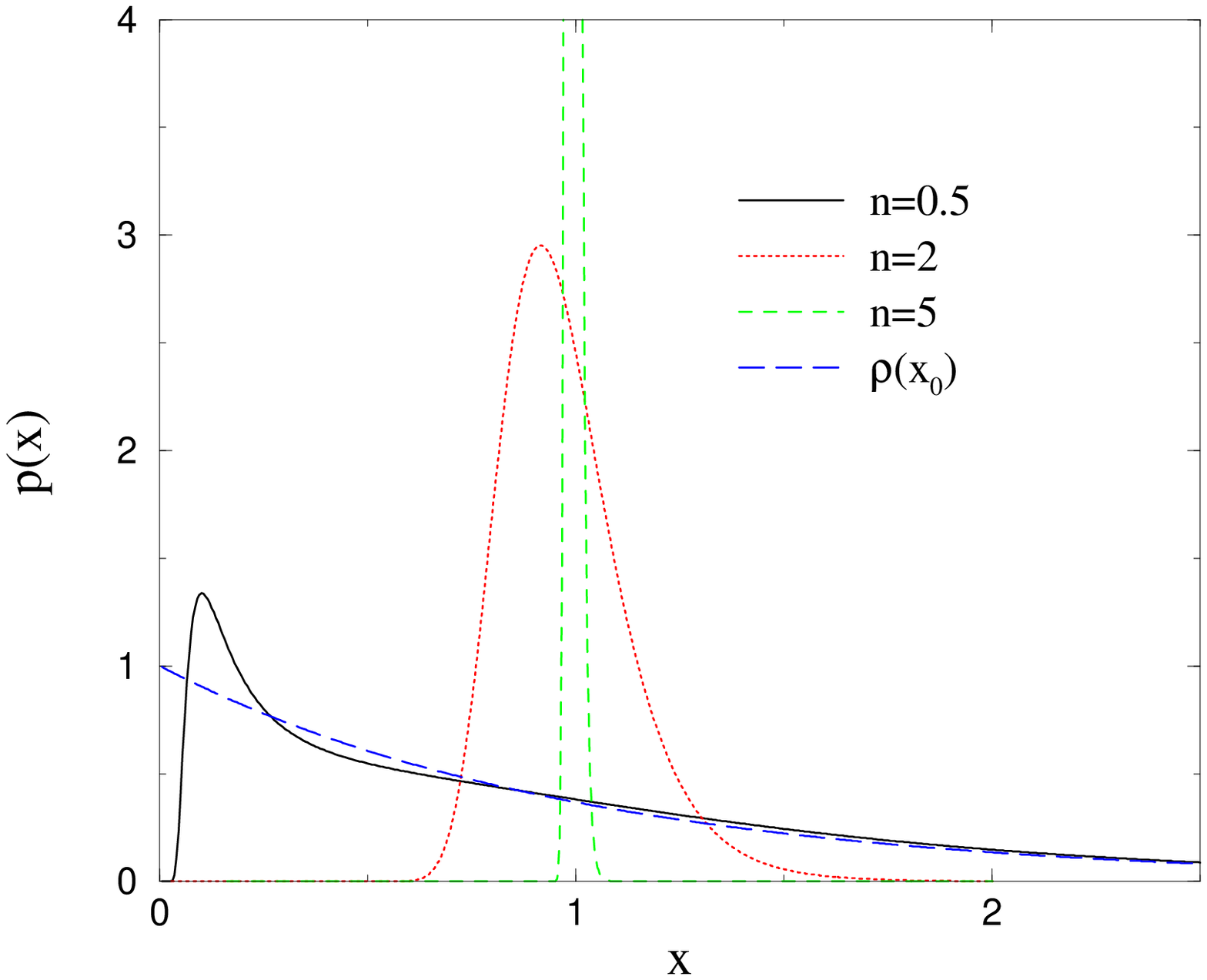} 
\caption{\label{pdf}Probability density functions of operation scales 
$s$ (left) and of consumptions $x$ (right) at equilibrium for 
$\epsilon=0.01$ and $n=0.5,~2$ and $5$.} 
\end{center} 
\end{figure}

As $n$ increases relative price fluctuations decrease. But the
decrease becomes very sharp close to $n_c$ for $\epsilon\ll 1$. In
this case, at $n_c$ price fluctuations suddenly drop to a level close
to zero. This is related to the behavior of the variable $x^*$ shown
in the right panel of Fig. \ref{sdpxdx}. In the region below $n_c$ the
average consumption level decreases. In this region firms take
advantage of the spread
\begin{equation} 
\frac{\avg{\delta x}}{\avg{x}}=\frac{\sqrt{\avg{x^2}-\avg{x}^2}}{\avg{x}} 
\end{equation} 
between scarce and abundant goods to make a living. But as $n$
approaches $n_c$, the spread in $x$ quickly drops to a very low value,
making life more difficult. Increasing $n$ beyond $n_c=2$, the economy becomes very selective toward increasingly efficient technologies which can perform the desired
transformation between commodities with a smaller decrease in the
average level $\avg{x}$ of consumption. This is clearly shown in
Fig. \ref{pdf} where we plot the probability densities of $x$ for
three different values of $n=0.5,~2$ and $5$. The right plot shows
that while for $n=0.5$ the distribution $P(x|\cdot)$, Eq.  \req{pdix},
retains the character of the distribution of initial endowments
$\rho(x_0)$, it becomes more and more peaked around $\avg{x_0}$ as $n$
increases. 
At the same time the distribution of $s$, Eq. \req{pdis},
becomes broader and broader.

The distribution of $s_i$, Eq. \req{pdis}, which may be considered as a proxy for firm sizes, gets broader and broader as $n$ increases. Interestingly mature economies, such as Japan~\cite{Takayasu} or the US~\cite{Axtell}, are characterized by a
very broad distribution of firm sizes, which we can put in relation
with $Q(s)$. The shape of the distribution found empirically is close
to a power law, which is different from \eqref{pdis}. However, it is
not difficult to derive a power law distribution of $s$ relaxing the
unrealistic assumption that all firms have the same value of $\Delta$
and $\epsilon$ \cite{physics_paper}.

\begin{figure}
\begin{center}
\includegraphics[width=8cm]{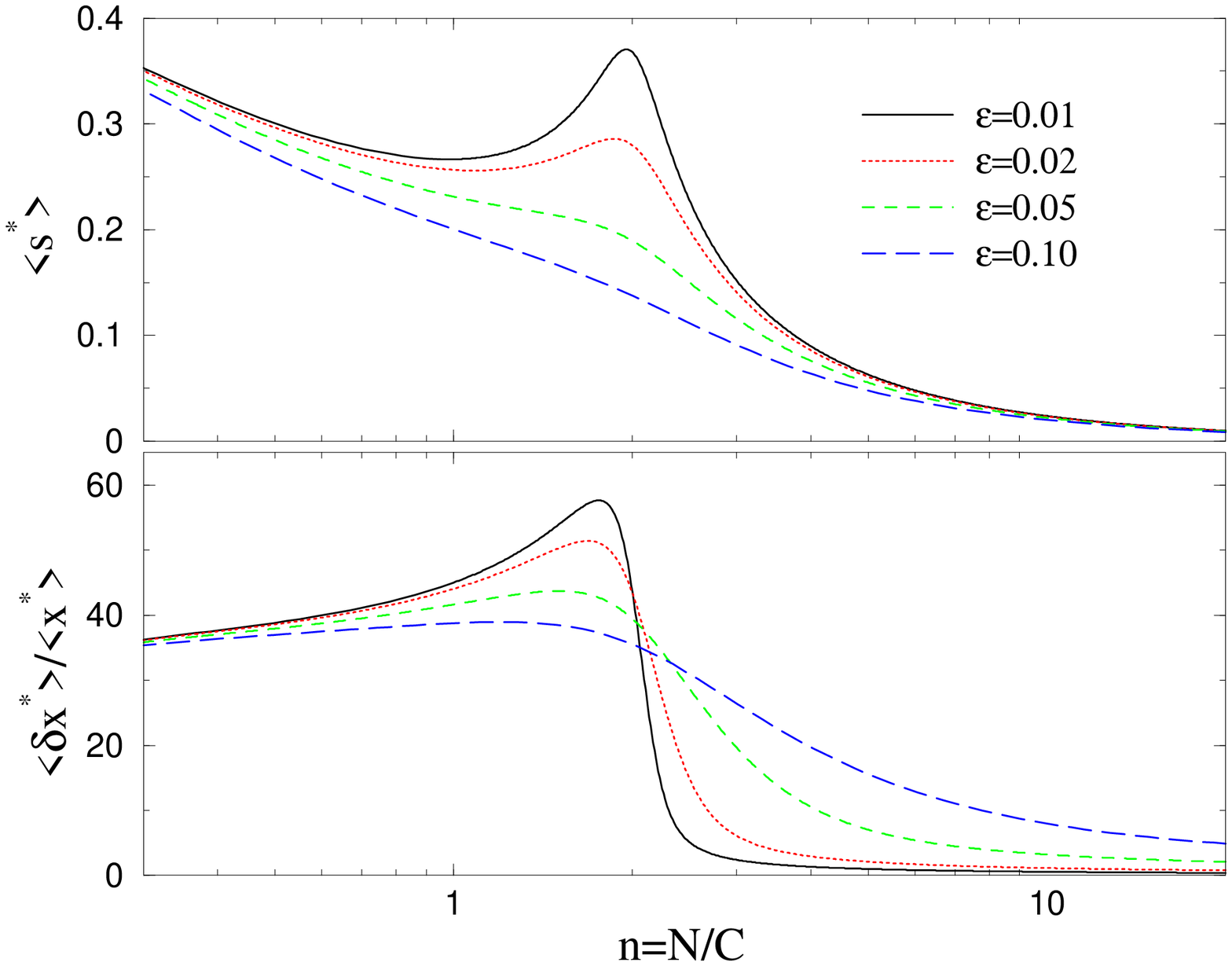}
\includegraphics[width=8.2cm]{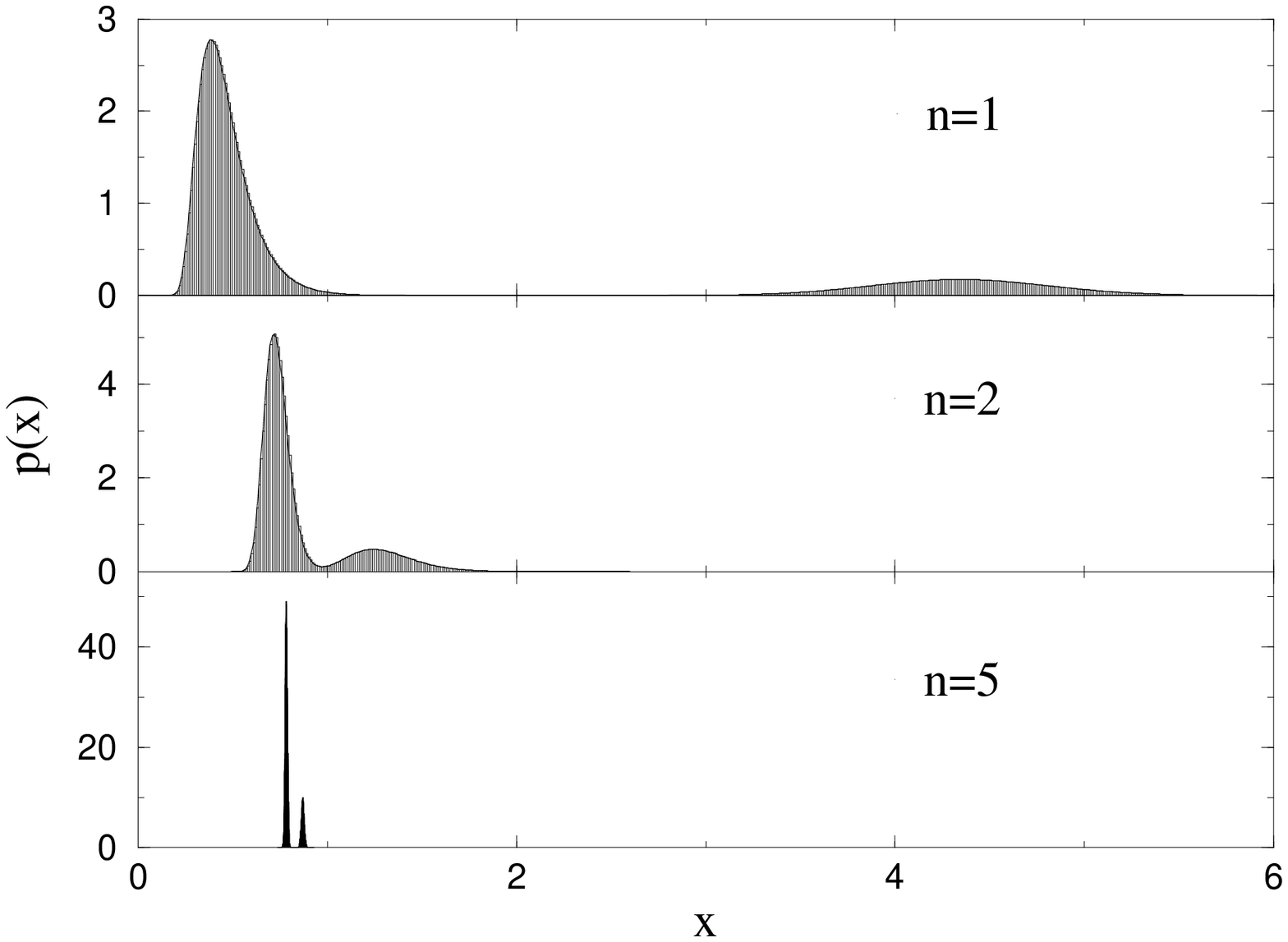}
\caption{\label{fmod}Left: Scale of production (top) and consumption
fluctuations (bottom) as a function of $n$ for a bimodal distribution
of initial endowments (Eq. \ref{deltamod}) with $f=0.2$ and $x_0=5$
($\epsilon=0.01$). Right: distribution of $x$ for $n=1,~2$ and $5$
from top to bottom.}
\end{center}
\end{figure}

The generic picture of the overall economy depicted thus far remains unchanged for different distributions $\rho(x_0)$ of initial endowments or for different utility functions $u(x)$. For example, Fig. 4 shows the results obtained with 
\begin{equation} \label{deltamod}
\rho(x_0)=(1-f)\delta(x_0)+f\delta(x_0-1). 
\end{equation} 
This captures the situation where only a fraction $f$ of the 
commodities is present in initial endowments (primary goods) whereas the remaining commodities have to be provided by the productive sector. 
The behavior of $\phi$, $\avg{x^*}$ and relative prices is very
similar to that found for the previous model. Fig. \ref{fmod} shows
that the average scale of production and the relative fluctuations of
$x^c$ show a qualitatively different behavior. Again the two regimes
with clearly distinct properties can be identified for $n<n_c$ and
$n>n_c$.

\section{Discussion}

The behavior of the solution with $n$ allows us to identify two classes of economies: mature economies ($n>2$) with a full-blown repertoire of technologies which closely saturate consumer's demand and immature economies ($n<2$) characterized by few technologies scattered in a large space of productive opportunities.

Strictly speaking, our considerations must be limited to comparative statics in view of the static nature of the equilibrium we study. However it is suggestive consider dynamic transitions between equilibria. In particular, a transitions $N\to N+1$, corresponding to the draw of a new technology, can be considered as the result of the discovery of a new method of combining inputs to produce desirable outputs; a new design as Romer \cite{Romer} calls it. Note however that in Romer's model innovation entails the discovery of a new intermediate commodity and there is no real ``combination" of inputs and no heterogeneity across technologies. By contrast, innovation in our model describes the expansion of the
frontier of feasible industrial transformation processes by discovery
of a new activities which is structurally different from existing ones\footnote{This is only one of the possible
modes of technological innovations. Innovations may also increase the efficiency of an existing technology, e.g. decreasing the input requirements, which may be captured by changes in $\epsilon_i$ and $\Delta_i$. Our focus here is on structural technological change.}; both the discrete nature of designs and the uncertainty of the discovery process are retained. 
Whether an innovation leading to the draw of a new technology is adopted or not will depends on the specific technologies that are already present\footnote{It has been argued that technological innovation is path dependent~\cite{Dosi}. This means that the draw of the $N+1^{\rm st}$ technology depends on the $N$ technologies which are already present. Such issues can clearly not be addressed within our quasi-static approach.}
Generalizing, one may also consider transitions $C\to C+1$ introducing a new commodity or characteristic \cite{Lancaster}, or which makes it possible to consume it.

\begin{figure}
\begin{center}
\includegraphics[width=10cm]{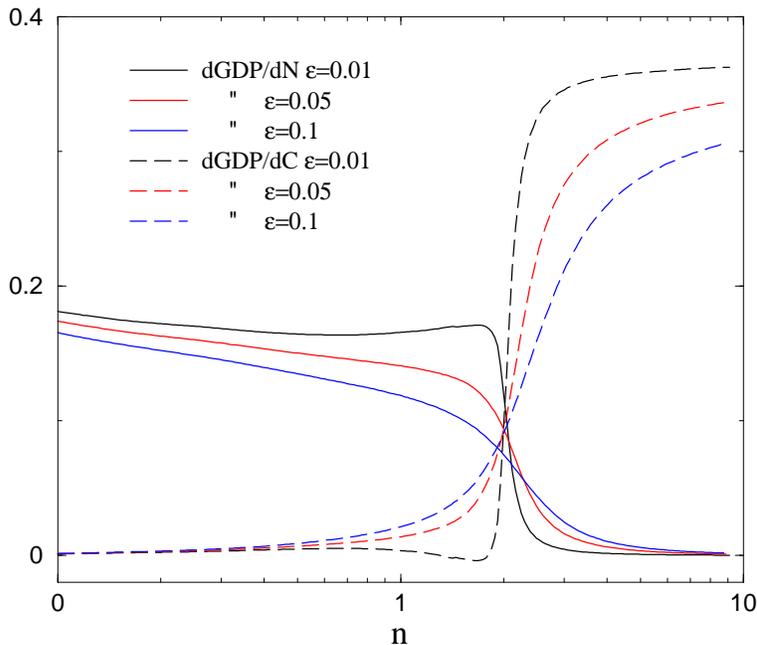}
\caption{\label{dgdp}Variation of GDP for $N\to N+1$ (full lines) and for $C\to C+1$ (dashed lines). Here $u(x)=\log x$ and $\rho(x_0)=e^{-x_0}$.}
\end{center}
\end{figure}

In such a dynamic view\footnote{For example consider the following cartoon of an evolving economy: In each period $t$ Nature endows the representative consumer with a bundle $\boldsymbol{x}_0(t)$ of commodities, drawn at random from the distribution $\rho_0$. Given the existing technologies, $\{\bfq_i:\,i=1,\ldots,N(t)\}$ this is transformed into the optimal bundle $\boldsymbol{x}(t)$ by the productive sector then $\boldsymbol{x}(t)$ is consumed at the end of period $t$. Finally an innovation event, i.e. a transition $N(t)\to N(t+1)=N(t)+1$ or $C(t)\to C(t+1)=C(t)+1$ may take place. It is implicit in this description that technological change occurs on a much longer time scale than that needed for the economy to reach equilibrium. This view may also describe the way in which existing technologies diffuse in a developing country. Then technologies are not discovered anew, but just become operative or feasible as e.g. human capital or infrastructure
accumulates, or as institutional constraints are removed.}, it is essential to consider the incentives for innovations in order to understand which transitions will most likely be generated endogenously. If we assume that transition rates depend on investments in research for new technologies, and that investment are related to expected changes in GDP generated by technological changes, then Fig. \ref{dgdp} suggests that the economy will drift towards $n\approx 2$. Indeed transitions $N\to N+1$ cause an increase in GDP which is sizeable for $n<2$ and almost negligible for $n>2$ (specially for small $\epsilon$). On the contrary, transitions $C\to C+1$ which decrease $n$ increase substantially GDP only for $n>2$. 

The same conclusion can be reached assuming that investment in research arise from the productive sector itself. Indeed the average scale $\avg{s^*}$ of activity of firms increases with $n$ for $n<2$, which means that a transition $N\to N+1$ causes an increase in the average scale of activities already active. 
This means that, at the equilibrium prices of the economy with $N<2C$ technologies, the profits of already existing firms increase on average when the new technology is introduced. Likewise, the decrease of $\avg{s^*}$ with $n$ for $n>2$ suggests that transitions $C\to C+1$ increase firms profits, on average, for $n>2$. This again yields a drift towards $n\approx 2$ due to endogenous technological change.

These arguments suggest that economies may spontaneously evolve towards a critical state, i.e. that they may be a further realization of self-organized criticality~\cite{SOC}. 

\section{Conclusions}

Summarizing, we have addressed the problem of calculating the general
equilibria of large linear production economies with random
technologies and a single consumer with tools of statistical
physics. In a nutshell, our results can be stated as follows. When the
ratio $n$ of the number of available technologies to the number of
commodities is below a threshold $n_c=2$, the average operation scale
grows as $n$ increases and roughly one-half of the firms are
active. For $n<2$, new technologies are easily accepted and the economy on the whole expands with $n$. When $n>n_c$, instead, the production sector is saturated, i.e. the number of active technologies converges to the number of commodities, and new
technologies are accepted only at the expense of reducing the
operation scales of the other technologies. The transition becomes
more and more sharp as the parameter $\epsilon$, measuring the inefficiency of each technology, approaches zero. From the consumer's viewpoint,
welfare increases with $n$ in both regimes. 
The main component of welfare increase with $n$ is different in the two regimes:
for $n<n_c$, welfare level grows with $n$ because the spread in
consumption levels $\avg{\delta x}$ decreases with the introduction of
technologies which transform abundant commodities into scarce
ones. For $n>n_c$, instead, growth arises from the introduction of
more efficient technologies, granting an increase in the level of
consumption $\avg{x^*}$. Accordingly, the relative spread of prices
$\avg{\delta p}/\avg{p}$ decreases with $n$.

Considering the incentives for technological innovations, we uncover a mechanism by which the economy self-organizes to the critical state $\approx 2$. Our model clearly is unrealistic in many respects. Still it may capture some novel aspects of structural technological change.
The extension of these approach to a fully dynamic setting, including capital accumulation, may shed new light on theories of endogenous economic growth.

Above all, we propose the use of statistical mechanics of disordered
systems to study the typical properties of the general equilibria of
large random economies. We have shown how these methods are able to
deal effectively with heterogeneity, providing a complete statistical
description of the equilibria, which is consistent with generic
results. The relevant quantities -- called order parameters -- are
naturally identified by the method. Given the non standard type of the
calculations involved, we also present computer experiments which
convincingly support our results.

The approach generalizes in a straightforward way to more complex
situations and we hope that this work will stimulate
cross-fertilization between the fields of economic theory and
statistical mechanics.

\section*{Acknowledgments} 

We are indebted to A. Bigano, D. Fiaschi, K.-G. Maler, F. Vega-Redondo, and M.A. Virasoro for useful discussions and to The Abdus Salam ICTP for hospitality and support within a thematic institute of the EU EXYSTENCE network of excellence. This work was supported in part by the European
Community's Human Potential Programme under contract
HPRN-CT-2002-00319, STIPCO. IPC thanks the Fund for Scientific
Research-Flanders, Belgium.

\appendix

\section*{Appendices} 

\section{The method}
\label{method} 
 
The standard technique to maximize a function of $N$ variables with 
$N\to\infty$ in statistical mechanics relies on the well-known 
steepest descent, or saddle point, method. Let $H_N(\cdot)$ be an 
extensive\footnote{I.e. such that there are two constants $k_+$ and 
$k_-$ satisfying $k_-N< H_N(\boldsymbol{s})<k_+N$ for all $N$ and 
$\boldsymbol{s}$.} function of $\boldsymbol{s}=\{s_i\}_{i=1}^N$, and 
imagine that we want to compute the maximum value of 
$H_N(\boldsymbol{s})/N$ in the limit $N\to\infty$.  Then 
\begin{equation}\label{sp2} 
\lim_{N\to\infty}\frac{1}{N}\max_{\boldsymbol{s}}~H_N(\boldsymbol{s}) 
=\lim_{\beta\to\infty}\lim_{N\to\infty}  
\frac{1}{\beta N}\log Z_N(\beta) 
\end{equation} 
where 
\begin{equation}\label{sp3} 
Z_N(\beta)=\int d\boldsymbol{s} ~e^{\beta H_N(\boldsymbol{s})} 
\end{equation} 
is called the {\em partition function} associated to $H_N$. Here $\int 
d\boldsymbol{s}$ stands for an $N$-dimensional integral on the whole 
domain of definition of $\boldsymbol{s}$. The idea of \req{sp3} is 
that the integral for $\beta\gg 1$ is dominated by regions where $H_N$ 
is maximal.  This recipe turns the problem of maximizing $h$ into that 
of calculating $Z_N$ and evaluating the asymptotic behavior of its 
logarithm. 
 
This task becomes much more difficult when $H_N$ depends on a set of 
random variables $\boldsymbol{q}$ with probability density 
$p(\boldsymbol{q})$. We denote this dependence by 
$H_N(\cdot|\boldsymbol{q})$.  The generic situation is that 
$\boldsymbol{q}$ enters the definition of the interactions among the 
$N$ components of $\boldsymbol{s}$ and $H_N$ is a sum over all 
interaction terms. In such situations, we expect that a sufficiently 
regular $H_N$ will obey the law of large numbers, so that e.g 
\begin{equation}\label{dfk} 
\lim_{N\to\infty} 
\frac{1}{N}\max_{\boldsymbol{s}}~H_N(\boldsymbol{s}|\boldsymbol{q})= 
\lim_{N\to\infty}\frac{1}{N}\avg{\left[\max_{\boldsymbol{s}}~ 
H_N(\boldsymbol{s}|\boldsymbol{q})\right]}_{\boldsymbol{q}} 
\end{equation} 
In other words, $\max H_N/N$ is expected to be a self-averaging 
quantity, namely to have vanishing sample-to-sample fluctuations in 
the limit $N\to\infty$. If one wanted to generalize ~\req{sp2} to the 
evaluation of \req{dfk}, one would have to compute the 
$\boldsymbol{q}$-average of the logarithm of the partition function 
$Z_N(\beta|\boldsymbol{q})$.  Unfortunately, the logarithm prevents 
every useful factorization of such an average and makes this way 
impracticable. 
 
The replica method is the standard statistical mechanical technique to 
circumvent this difficulty. Using the formula 
\begin{equation}\label{replica} 
\log Z_N(\beta|\boldsymbol{q})=\lim_{r\to 
0}\frac{[Z_N(\beta|\boldsymbol{q})]^r-1}{r} 
\end{equation} 
we can reduce our problem to that of computing 
$\avgq{[Z_N(\beta|\boldsymbol{q})]^r}$.  This is feasible for integer 
values of $r$ because it amounts to computing 
\begin{equation}\label{zr} 
[Z_N(\beta|\boldsymbol{q})]^r=\l[\int e^{\beta 
H_N(\boldsymbol{s}|\boldsymbol{q})}d\boldsymbol{s}\r]^r =\int 
e^{\beta\sum_{a=1}^r H_N(\boldsymbol{s}_a|\boldsymbol{q})} 
\prod_{a=1}^r d\boldsymbol{s}_a 
\end{equation} 
which is the partition function of $r$ ``replicas'' of the original 
system \emph{with the same disorder realization $\boldsymbol{q}$} 
(hence the name of the method). The last step consists in performing 
an analytic continuation for real values of $r$ and taking the limit 
$r\to 0$: 
\begin{equation}\label{repl} 
\lim_{N\to\infty}\frac{1}{N}\avgq{\max_{\boldsymbol{s}}~ 
H_N(\boldsymbol{s}|\boldsymbol{q})}=\lim_{N\to\infty}\lim_{\beta\to\infty} 
\lim_{r\to 0}\frac{1}{\beta N 
r}\log \avgq{[Z_N(\beta|\boldsymbol{q})]^r}. 
\end{equation} 
The existence and uniqueness of the limit $r\to 0$, which looks
somewhat bizarre, have been much debated in the physics literature
(see \cite{MPV} for a discussion).  Even if this method remains a
formally non-rigorous procedure, several rigorous mathematical results
confirm its validity in problems that are more complex that the one we
deal with here \cite{talagrandH,talagrandSG}. We hope this (together
with the agreement with computer experiments) gives the reader a
sufficient level of confidence to accept the $r\to 0$ passage.
 
The technical part of the calculation lies in the introduction of a 
finite number of auxiliary integration variables $\boldsymbol{\theta}= 
\{\theta_1,\ldots,\theta_k\}$ allowing the averaged replicated 
partition function to be re-cast in the form 
\begin{equation}\label{zzzeta} 
\avgq{[Z_N(\beta|\boldsymbol{q})]^r}\simeq 
\int e^{\beta N r [h(\boldsymbol{\theta})+o(r,\beta,\boldsymbol{\theta})]} 
d\boldsymbol{\theta} 
\end{equation} 
where $h(\cdot)$ is some function and $o(r,\beta,\cdot)\to 0$ in the
limits $\beta\to \infty$, $r\to 0$. The $\boldsymbol{\theta}$
variables are called {\em order parameters}. Their nature and number
are dictated by the mathematical structure of the problem (see
Appendix \ref{appconti} for the details of our case). Finally,
assuming that the limits $r\to 0$ and $N\to\infty$ commute, the latter
can be taken first in (\ref{repl}) thus allowing to evaluate
(\ref{zzzeta}) by the saddle-point method as
\begin{equation}\label{saddle} 
\avgq{[Z_N(\beta|\boldsymbol{q})]^r} 
\sim e^{\beta N r 
  [h(\boldsymbol{\theta}^*)+o(r,\beta,\boldsymbol{\theta}^*)]}  
\end{equation} 
where $\boldsymbol{\theta}^*$ is the saddle point value of 
$\boldsymbol{\theta}$ which dominates the integral in 
\req{zzzeta}. Hence, putting things together, 
\begin{equation}\label{ghtre} 
\lim_{N\to\infty}\frac{1}{N}\avgq{\max_{\boldsymbol{s}}~ 
H_N(\boldsymbol{s}|\boldsymbol{q})}= 
\lim_{N\to\infty}\lim_{\beta\to\infty}  
\lim_{r\to 0}\frac{1}{\beta N r}\log~ e^{\beta 
N r [h(\boldsymbol{\theta}^*)+o(r,\beta,\boldsymbol{\theta}^*)]} 
= h(\boldsymbol{\theta}^*) 
\end{equation} 
 
The core of the procedure lies in \req{zzzeta} where, by a lengthy
calculation one identifies the relevant order parameters
$\boldsymbol{\theta}$ and the function $h$.  This crucial but
technical step is presented below (appendix \ref{appconti}) for our
problem.

\section{The explicit calculation of the representative agent problem} 
 \label{appconti}
 
The partition function in our case reads 
\begin{equation} 
 Z_N(\beta|\boldsymbol{q})=\int_0^\infty e^{\beta 
U\big(\boldsymbol{x}_0 
+\sum_{i=1}^Ns_i\boldsymbol{q}_{i}\big)}d\boldsymbol{s} 
\label{partition_function} 
\end{equation} 
with $U(\boldsymbol{x})$ the utility function of the representative
consumer. As stated above, in order to analyze the statistical
properties of the equilibria we have to evaluate
$\avgq{[Z_N(\beta|\boldsymbol{q})]^r}$ and resort to \req{repl}, with
$H_N$ given in our case by $U$ and with all the necessary
constraints. Before proceeding, we shall introduce some useful
definitions and identities. The first one is the $\delta$-function
$\delta(x)$ which is defined through the relation
\begin{equation} 
f(y)=\int_{\mathbb{R}} \delta(x-y)f(x)dx 
\label{delta} 
\end{equation} 
for any function $f(\cdot)$ and $y\in\mathbb{R}$. We will also use the 
exponential representation of the $\delta$-function 
\begin{equation} 
\delta(x)=\int_{\mathbb{R}} e^{\ii\widehat{x}x}~\frac{d\widehat{x}}{2\pi} 
\label{delta_repre} 
\end{equation}
Another mathematical tool we will use is the Gaussian or 
Hubbard-Stratonovich transformation, viz. 
\begin{equation} 
\exp\l[\frac{b^2}{2}\r]=\int_{\mathbb{R}} 
\exp\l[-\frac{x^2}{2}+bx\r]~\frac{dx}{\sqrt{2\pi}} 
\end{equation} 
which allows to linearize arguments of exponentials at the cost of 
introducing averages over Gaussian r.v.'s. Now, in order to perform 
our calculation it is convenient to replace the consumption variables 
$\boldsymbol{x}$ by writing explicitly the market clearing condition 
\eqref{mc} in the partition function \eqref{partition_function}. To do 
so we use the defining property \eqref{delta} of 
$\delta$-distributions and write 
\begin{equation} 
Z_N(\beta|\boldsymbol{q})=\int_{0}^\infty d\boldsymbol{x} 
\int_{0}^\infty d\boldsymbol{s}~ e^{\beta U(\boldsymbol{x})} 
\prod_{c=1}^C\delta\Big(x^c-x^c_0-\sum_{i=1}^Ns_iq_{i}^c\Big) 
\end{equation} 
As already explained in Sec. \ref{method} we will have to take the
following steps: (a) average the partition function of $r$ replicas
over technologies, as in \eqref{zzzeta}; (b) identify the correct
order parameters of the problem to write the latter average as in
\req{zzzeta}; (c) take the limits $N\to\infty$ and $r\to 0$ and get
something of the form of \req{ghtre}; and finally (d) find the values
of the order parameters at the competitive equilibrium (i.e. when
$\beta\to\infty$).
 
The partition function of $r$ replicas reads 
\begin{equation}\label{pfff} 
[Z_N(\beta|\boldsymbol{q})]^r=\int_{0}^\infty \prod_{a=1}^r 
d\boldsymbol{x}_a \int_{0}^\infty \prod_{a=1}^r d\boldsymbol{s}_a 
~e^{\beta\sum_{a=1}^r U(\boldsymbol{x}_a)} 
\prod_{a=1}^r\prod_{c=1}^C\delta\Big(x^c_a-x^c_0-\sum_{i=1}^N 
s_{i,a}q_{i}^c\Big) 
\end{equation} 
Notice that the dependence on the technologies appears in the market
clearing condition only, so that the average $\avgq{\ldots}$ involves
only the last part of $[Z_N(\beta|\boldsymbol{q})]^r$. This average
one must take into account the constraint \eqref{constraint}, i.e.
\begin{equation} 
\avgq{\cdots}=\frac{\avgq{\prod_{i=1}^N\delta\big(\sum_{c=1}^C 
 q_i^c+\epsilon\big) 
 (\cdots)}'}{\avgq{\prod_{i=1}^N\delta\big(\sum_{c=1}^C 
 q_i^c+\epsilon\big)}'} 
\end{equation} 
where $\avgq{\cdots}'$ stands for the average over unconstrained
i.i.d. Gaussian vectors $\bfq$ with zero mean and variance
$\avgq{\bfq^2}=\sum_c\avgq{(q^c)^2}=\Delta$. Using Eq.
\eqref{delta_repre} for the constraints, the denominator becomes
\begin{equation} 
\avgq{\prod_{i=1}^N\delta\big(\sum_{c=1}^C q_i^c+\epsilon\big)}'=
\prod_{i=1}^N\frac{1}{\sqrt{2\pi\Delta}}\exp\Big[
-\frac{\epsilon^2}{2\Delta}\Big]
\end{equation} 
while for the numerator we get 
\begin{multline} 
\avgq{\prod_{i=1}^N\delta\big(\sum_{c=1}^C q_i^c+\epsilon\big) 
\prod_{a=1}^r\prod_{c=1}^C\delta\Big(x^c_a-x^c_0-\sum_{i=1}^N 
s_{i,a}q_{i}^c\Big)}'=\\= \int \prod_{i=1}^N 
\frac{d\widehat{z}_i}{2\pi}\int \prod_{a=1}^r 
\frac{d\widehat{\boldsymbol{x}}_a}{2\pi}\exp\Big[\ii\epsilon\sum_{i=1}^N\widehat{z}_i 
+\ii\sum_{a=1}^r\sum_{c=1}^C\widehat{x}_a^c\big(x^c_a-x^c_0\big) 
-\frac{\Delta}{2C}\sum_{i=1}^N \sum_{c=1}^C\Big(\widehat{z}_i 
-\sum_{a=1}^r\widehat{x}_a^c s_{i,a}\Big)^2\Big] 
\end{multline} 
Notice that the expected values involved in these calculations are all
of the form $\psi(y)=\avgq{e^{iyq_i^c}}$. This is the characteristic
function of $q_i^c$ and for the assumed Gaussian distribution it takes
the form $\psi(y)=e^{-\Delta y^2/(2 C)}$. This result can however be
extended to any distribution of $q_i^c$ with
$\psi(y)=\tilde\psi(y/\sqrt{C})$ with a leading behavior
$\tilde\psi(x)=-\Delta x^2/2+O(x^3)$. Indeed all higher order terms in
the power expansion of $\tilde\psi$ give vanishingly small
contributions with respect to the first, in the limit $C\to\infty$.

Gathering all the terms, we have 
\begin{gather} 
\avgq{[Z_N(\beta|\boldsymbol{q})]^r}=\int \prod_{i=1}^N 
\frac{d\widehat{z}_i}{2\pi} \int 
\prod_{a=1}^r\frac{d\widehat{\boldsymbol{x}}_a}{2\pi} \int_{0}^\infty 
\prod_{a=1}^r d\boldsymbol{x}_a\int_{0}^\infty \prod_{a=1}^r 
d\boldsymbol{s}_a \exp\Big[\beta\sum_{a=1}^r 
U(\boldsymbol{x}_a)+\ii\epsilon\sum_{i=1}^N\widehat{z}_i+\\ 
+\ii\sum_{a=1}^r\sum_{c=1}^C\widehat{x}_a^c\big(x^c_a-x^c_0\big) 
-\frac{\Delta}{2C}\sum_{i=1}^N \sum_{c=1}^C\Big(\widehat{z}_i 
-\sum_{a=1}^r\widehat{x}_a^c s_{i,a}\Big)^2\Big] 
\Bigg[\prod_{i=1}^N\frac{1}{\sqrt{2\pi\Delta}} 
\exp\Big[-\frac{\epsilon^2}{2\Delta}\Big]\Bigg]^{-1} \label{mixx}
\end{gather} 
In order to write the above in a form as simple as \eqref{zzzeta}, the
set of order parameters to be introduced must allow for a decoupling
of the integrals over the variables $\widehat{z}_i$, $s_{i,a}$ and
$\widehat{ x}_a^c$ in such a way that the integrals on the different
variables can be factorized.  Here it is enough to introduce the
following order parameters
\begin{equation} 
\omega_{ab}=\frac{1}{N}\sum_{i=1}^Ns_{i,a}s_{i,b} 
\qquad\text{and}\qquad k_a=\frac{1}{N}\sum_{i=1}^N\widehat{z}_is_{i,a} 
\end{equation} 
through identities such as
\begin{equation}\label{identity}
1=\int
d\omega_{ab}~ N\delta\left(N\omega_{ab}-\sum_{i=1}^Ns_{i,a}s_{i,b}\right)=
\int \frac{d\omega_{ab}d\widehat{\omega}_{ab}}{2\pi
\ii/N}e^{\widehat{\omega}_{ab}[N\omega_{ab}-\sum_is_{i,a}s_{i,b}]}
\end{equation}
Then last term in the exponent of the numerator of Eq. \eqref{mixx} becomes
\begin{equation}
\sum_{i=1}^N\Big(\widehat{z}_i 
-\sum_{a=1}^r\widehat{x}_a^c s_{i,a}\Big)^2 =\sum_{i=1}^N \widehat{z}_i^2-2N\sum_{a=1}^r k_a\widehat{x}_a^c+N\sum_{a,b=1}^r\omega_{ab}\widehat{x}_a^c\widehat{x}_b^c
\end{equation}
This allows us to separate the problem into three parts. Indeed we can
re-cast the replicated partition function in the form of a set of
integrals over the order parameters:
\begin{equation} 
\avgq{[Z_N(\beta|\boldsymbol{q})]^r}= 
\int\prod_{a,b=1}^r\frac{d\omega_{ab}d\widehat{\omega}_{ab}}{4\pi 
\ii/N} \int\prod_{a=1}^r \frac{dk_a d\widehat{k}_a}{2\pi \ii/N} 
\exp\big[N 
h(\{\omega_{ab}\},\{\widehat{\omega}_{ab}\},\{k_a\},\{\widehat{k}_a\})\big] 
\label{arpf} 
\end{equation} 
where $h=g_1+g_2+g_3$ is the sum of three terms which can be computed
independently. In particular
\begin{gather} 
g_1\equiv
g_1(\{\omega_{ab}\},\{\widehat{\omega}_{ab}\},\{k_a\},\{\widehat{k}_a\})=
-\frac{1}{2}\sum_{a,b=1}^r\widehat{\omega}_{ab}\omega_{ab}-\sum_{a=1}^r
\widehat{k}_{a} k_a\\ g_2\equiv
g_2(\{\widehat{\omega}_{ab}\},\{\widehat{k}_a\})=\log\int
\frac{d\widehat{z}}{2\pi} \int_{0}^\infty \prod_{a=1}^r d
s_a\exp\Big[\frac{1}{2}\sum_{a,b=1}^r\widehat{\omega}_{ab}s_as_b
+\widehat{z}\sum_{a=1}^r\widehat{k}_{a}s_a+\ii\epsilon\widehat{z}
-\frac{\Delta}{2}\widehat{z}^2\Big]\nonumber\\
-\log\Bigg[\frac{1}{\sqrt{2\pi\Delta}}\exp\Big[-\frac{\epsilon^2}{2\Delta}\Big]\Bigg]\\
g_3\equiv
g_3(\{\omega_{ab}\},\{k_a\})=\frac{1}{N}\sum_{c=1}^C\log\int\prod_{a=1}^r
\frac{d\widehat{x}_a}{2\pi} \int_{0}^\infty \prod_{a=1}^r d
x_a\exp\Big[\beta\sum_{a=1}^r u(x_a)
+\ii\sum_{a=1}^r\widehat{x}_a\big(x_a-x^c_0\big)\nonumber\\
-\frac{n\Delta}{2}\sum_{a,b=1}^r\widehat{x}_a\widehat{x}_b \omega_{ab}
+n\Delta \sum_{a=1}^r\widehat{x}_a k_a\Big]
\end{gather} 
with $n=N/C$. The order parameters $\widehat{k}_a$ have appeared after
using an identity similar to \eqref{identity} for $k_a$.  Now
\req{arpf} is precisely of the form \req{zzzeta}.
 
In the limit $N\to\infty$ the integrals appearing in \eqref{arpf} are
dominated by the contributions coming from the saddle-point of $h$ and
the solution of our specific problem can be written as
\begin{equation}\label{fgr} 
\lim_{N\to\infty}^{(n)}\frac{1}{N}~\avgq{\max_{\boldsymbol{s}}~
U(\boldsymbol{x})} =\lim_{\beta\to\infty} \lim_{r\to 0}\frac{1}{\beta
r}~
h(\{\omega^*_{ab}\},\{\widehat{\omega}^*_{ab}\},\{k^*_a\},\{\widehat{k}^*_a\})
\end{equation} 
where the ${}^*$ means that parameters take their saddle point value,
i.e. those who solve the system of equations
\begin{equation}\label{saddlep}
\frac{\partial h}{\partial \omega_{ab}}=0,~~\frac{\partial h}{\partial
\widehat{\omega}_{ab}}=0,~~\frac{\partial h}{\partial
k_{a}}=0,~~\frac{\partial h}{\partial \widehat{k}_{a}}=0
\end{equation}
for all $a,b=1,\ldots,r$. Ideally one should first solve these
equations for generic $r$ and then take the limit $r\to 0$.

A word about the meaning of the order parameters introduced thus far
is in order before taking the limit $r\to 0$. Indeed, $\omega_{ab}$ is
an $r\times r$ matrix for integer $r$, but it is not clear how can we
handle it in the limit $r\to 0$. When we replicated the partition
function passing from \req{partition_function} to \req{pfff} we
essentially passed from a problem in which $U(\boldsymbol{x}_a)$ is to
be maximized to an equivalent problem in which $\sum_a
U(\boldsymbol{x}_a)$ is to be maximized. The latter sum is evidently
left unchanged by a permutation of the replica indexes
$1,\ldots,r$. Hence it must be expected that, as long as there is a
unique maximum (as in this case), replica permutation symmetry is
preserved also by the solution of Eq.s \eqref{saddlep}. Then we expect
a solution of the form
\begin{gather} 
\omega_{ab}^*=\Omega\delta_{ab}+\omega(1-\delta_{ab})\nonumber\\ 
\widehat{\omega}_{ab}^*=\widehat{\Omega}\delta_{ab}+\widehat{\omega} 
(1-\delta_{ab}) 
\label{rsaaaa}\\ k_a^*=k\qquad \widehat{k}_a^*=\widehat{k}\nonumber 
\end{gather} 
This is the so-called replica-symmetric Ansatz, which simply expresses
the conservation of the permutation symmetry. When multiple maxima
with different statistical properties exist, this Ansatz fails because
replicas can converge to maxima with different properties, and hence
replicas are no more equivalent. This situation is ruled out in our
case by the nature of the function we want to maximize.

With Eq.s \eqref{rsaaaa}, it is easy to find an analytic expression of
the function $g_1$, $g_2$ and $g_3$ in terms of $r$ and to perform the
limit $r\to 0$. Substituting (\ref{rsaaaa}) into the definitions of
$g_1$, $g_2$ and $g_3$, after some straightforward algebraic
manipulations one finds
\begin{gather} 
\lim_{r\to0}\frac{1}{r}g_1=-\frac{1}{2}\Big(\widehat{\Omega}\Omega- 
\widehat{\omega}\omega\Big) -\widehat{k} k\\ 
\lim_{r\to0}\frac{1}{r}g_2=\left\langle\log\int_0^\infty ds ~e^{\beta 
V(s;t)}\right\rangle_{t}\\ \lim_{r\to0}\frac{1}{r}g_3=\frac{1}{n} 
\left\langle\log \int_0^\infty dx~e^{\beta 
W(x;t,x_0)}\right\rangle_{t,x_0} 
\end{gather} 
where 
\begin{gather} 
\beta W(x;t,x_0)\equiv\beta u(x)-\frac{\Big(x-x_0+\sqrt{n\Delta 
\omega} t -\ii n\Delta 
k\Big)^2}{2n\Delta(\Omega-\omega)}-\frac{1}{2}\log[2\pi 
n\Delta(\Omega-\omega)]\\ \beta 
V(s;t)\equiv\frac{\widehat{\Omega}-\widehat{\omega}}{2}s^2+\Big[t\Big( 
\frac{\widehat{k}^2 }{\Delta}+\widehat{\omega}\Big)^{1/2}+ 
\ii\widehat{k}\frac{\epsilon}{\Delta}\Big]s 
\end{gather} 
 
We must finally evaluate the limit $\beta\to \infty$.  In this limit, 
a somewhat special role is played by the quantity 
$\chi=\beta(\Omega-\omega)$. Notice that 
\begin{equation}
\Omega-\omega=\frac{1}{2N}\sum_{i=1}^N(s_{i,a}-s_{i,b})^2
\end{equation}
is the distance between two replicas. The two vectors
$\boldsymbol{s}_a$ and $\boldsymbol{s}_b$ both converge to the unique
solution of the maximization problem as $\beta\to \infty$. Hence, we
also expect the distance $\Omega-\omega$ to vanish in this limit. But
looking e.g. at $W/\beta$ one realizes that in order to avoid annoying
divergences or trivial limits this quantity must vanish in such a way
that the product $\beta(\Omega-\omega)$ stays finite. In other terms,
one wants that $\Omega-\omega\sim 1/\beta$ for large $\beta$. If this
is the case, the maximization problem has a well-defined
solution. Hence we assume that $\lim_{\beta\to\infty}\chi$ is
finite. Similar arguments lead to the introduction of the following
re-defined order parameters, which remain finite as $\beta\to\infty$:
\begin{gather} 
\chi= n\Delta\beta(\Omega-\omega),\qquad 
\widehat{\chi}=-\frac{\widehat{\Omega}-\widehat{\omega}}{\beta}, 
\qquad \kappa=-\ii n\Delta k\\ 
\widehat{\kappa}=\ii\frac{\widehat{k}}{\Delta \beta}, \qquad 
\widehat{\gamma}=\beta^{-2}\widehat{\omega}
\end{gather} 
Inserting these in the previous expressions we find that the r.h.s. of
(\ref{fgr}) (which we for simplicity denote again by $h$) can
be written as
\begin{equation} 
h(\Omega,\kappa,\widehat{\kappa},\widehat{\gamma},\chi,\widehat{\chi})= 
\frac{1}{2}\Big(\Omega\widehat{\chi}-\frac{\widehat{\gamma}\chi}{n\Delta}\Big) 
-\frac{1}{n}\widehat{\kappa} \kappa + 
\frac{1}{\beta}\left\langle\log\int_0^\infty ds~ e^{\beta 
V(s;t)}\right\rangle_{t} +\frac{1}{n \beta}\left\langle\log 
\int_0^\infty dx ~e^{\beta W(x;t,x_0)}\right\rangle_{t,x_0} 
\end{equation} 
where now the functions $V$ and $W$ read 
\begin{gather} 
 W(x;t,x_0)= u(x)-\frac{\big(x-x_0+ \kappa+\sqrt{n\Delta \Omega} t\big)^2}{2\chi} 
\\ V(s;t)=-\frac{\widehat{\chi}}{2}s^2+\big(t\sqrt{\widehat{\gamma} 
-\Delta\widehat{\kappa}^2}+\widehat{\kappa}\epsilon \big)s 
\end{gather} 
We neglected the last term in $W$ because it is vanishingly small in
the limit $\beta\to\infty$ when $\chi$ is finite.
 
When $\beta\to\infty$, again by steepest descent reasoning, only the 
maxima of $V$ and $W$ contribute to the integrals over $s$ and $x$. 
Therefore we can write the final expression for $h$ as 
\begin{multline} 
h(\Omega,\kappa,\widehat{\kappa},\widehat{\gamma},\chi,\widehat{\chi})= 
 \left\langle\max_{s\geq 0}\Big[-\frac{\widehat{\chi}}{2}s^2+ 
 \big(t\sqrt{\widehat{\gamma}-\Delta\widehat{\kappa}^2}+ 
 \widehat{\kappa}\epsilon \big)s 
 \Big]\right\rangle_{t}+\frac{1}{2}\Big(\Omega\widehat{\chi}-\frac{\widehat{\gamma}\chi}{n\Delta}\Big) 
 -\frac{1}{n}\widehat{\kappa} \kappa+\\ +\frac{1}{n 
 }\left\langle\max_{x\geq 0}\Big[u(x) -\frac{\big(x-x_0+ 
 \kappa+\sqrt{n\Delta \Omega} t\big)^2}{2\chi}\Big] \right\rangle_{t,x_0} 
\label{h1} 
\end{multline} 
The difference between this expression and the one appearing in
\eqref{h} is again a trivial redefinition of the order parameters.  If
we let now $x^*(t,x_0)$ and $s^*(t)$ be the values maximizing the
functions $W$ and $V$, respectively, and therefore given by
\eqref{sstar} and \eqref{xstar}, we can then expand \eqref{h1} to
obtain
\begin{multline} 
h(\Omega,\kappa,\widehat{\kappa},\widehat{\gamma},\chi,\widehat{\chi})= 
-\frac{\widehat{\chi}}{2}\left\langle(s^*)^2\right\rangle_{t}+\sqrt{\widehat{\gamma} 
-\Delta\widehat{\kappa}^2}\left\langle ts^*\right\rangle_{t}+ 
\widehat{\kappa}\epsilon\left\langle s^* \right\rangle_{t}+ 
\frac{1}{2}\Big(\Omega\widehat{\chi}-\frac{\widehat{\gamma}\chi}{n\Delta}\Big) 
-\frac{1}{n}\widehat{\kappa} \kappa+\\ +\frac{1}{n }\left\langle 
u(x^*)\right\rangle_{t,x_0} -\frac{1}{2n\chi 
}\left\langle\big(x^*-x_0+ \kappa +\sqrt{n\Delta \Omega} t\big)^2 
\right\rangle_{t,x_0} 
\end{multline} 
 The last step is to derive the saddle-point equations from which the 
values that the order parameters take on at equilibrium can be 
calculated. Computing the derivatives of $h$ with respect to the order 
parameters we get 
\begin{eqnarray} 
\frac{\partial h}{\partial \Omega}&=&\frac{1}{2}\widehat{\chi} 
-\frac{1}{2\chi }\sqrt{\frac{\Delta}{n\Omega}}\left\langle\big(x^*-x_0+ 
 \kappa+\sqrt{n\Delta \Omega} t\big)t \right\rangle_{t,x_0}\\ 
\frac{\partial h}{\partial \kappa}&=&-\frac{1}{n}\widehat{\kappa} 
-\frac{1}{n\chi }\left\langle x^*-x_0+ \kappa+ 
\sqrt{n\Delta \Omega} t \right\rangle_{t,x_0}\\ 
\frac{\partial h}{\partial \widehat{\kappa}}&=& 
\frac{-\Delta\widehat{\kappa}}{\sqrt{\widehat{\gamma}-\Delta\widehat{\kappa}^2}} 
\left\langle ts^*\right\rangle_{t}+\epsilon\left\langle s^* \right\rangle_{t}-\frac{1}{n} \kappa\\ 
\frac{\partial h}{\partial \widehat{\gamma}}&=&\frac{1}{2\sqrt{\widehat{\gamma} 
-\Delta\widehat{\kappa}^2}}\left\langle ts^*\right\rangle_{t}-\frac{\chi}{2n\Delta}\\ 
\frac{\partial h}{\partial \chi}&=&-\frac{\widehat{\gamma}}{2n\Delta} 
+\frac{1}{2n\chi^2 }\left\langle\big(x^*-x_0+ 
 \kappa+\sqrt{n\Delta \Omega} t\big)^2 \right\rangle_{t,x_0}\\ 
\frac{\partial h}{\partial \widehat{\chi}}&=& 
-\frac{1}{2}\left\langle(s^*)^2\right\rangle_{t}+\frac{1}{2}\Omega 
\end{eqnarray} 
Using the relation \eqref{xstar} and setting $p=-\widehat{\kappa}$, $\sigma=\sqrt{\widehat{\gamma}-\Delta\widehat{\kappa}^2}$ 
we finally arrive at Eq.s (\ref{p}--\ref{k}).

\section{The p.d.f.'s of $s$ and $x$} 
\label{apppdf}
 
We illustrate here the procedure for calculating the conditional
probability density of $x$ (the equilibrium consumption) given $x_0$
(the initial endowment). The derivation of the distribution of $s$
follows exactly the same lines. One can start from the identity
\begin{equation} 
P(x|x_0)=\int_{-\infty}^\infty\frac{dt}{\sqrt{2\pi}}~e^{-t^2/2} 
\delta[x-x^*(t,x_0)]  
\end{equation} 
Then one can make use of the property 
$\delta(x-x^*)=|f'(x^*)|\delta[f(x)]$, where $f(x)$ is a function with 
an unique root in $x^*$. From \req{xstar}, we take  
\begin{equation} 
f(x)=\frac{x-x_0-\chi u'(x)+k}{\sqrt{n\Delta \Omega}}+t  
\end{equation} 
so that 
\begin{equation} 
P(x|x_0)=\int_{-\infty}^\infty\frac{dt}{\sqrt{2\pi}}~e^{-t^2/2} 
\frac{1-\chi u''(x)}{\sqrt{n\Delta \Omega}}\delta\left[t+\frac{x-x_0-\chi 
u'(x)+k} {\sqrt{n\Delta \Omega}}\right]  
\end{equation} 
From this, taking the integral over $t$, one immediately finds 
\req{pdix}. 
 
\section{Calculation of $h$ at the saddle point and derivation of Walras' law} 
\label{appgeneric}
 
Substituting $s$ with $s^*(t)$ \req{sstar} and $x$ with $x^*(t,x_0)$ 
\req{xstar} we can re-write $h$ as 
\begin{multline} 
h=\sigma\avg{s^*t}_t-\epsilon\avg{s^*}_tp-\frac{1}{2}\hat\chi\avg{(s^*)^2}_t 
+\frac{1}{2}\hat\chi \Omega 
+\frac{kp}{n}-\frac{\chi\sigma^2}{2n\Delta}-\frac{\chi p^2}{2n}+\\+ 
\frac{1}{n}\avg{u(x^*)}_{t,x_0} 
-\frac{1}{2n\chi}\avg{(x^*-x_0+k+\sqrt{n\Delta \Omega}t)^2}_{t,x_0}  
\end{multline} 
Now it's a simple algebraic problem. For the first term on the 
r.h.s. we use \req{chi}; for the second and the fifth we use \req{k}; 
the third and fourth cancel because of \req{Omega}; finally for the 
last term we use \req{xstar} and then \req{sigma} to find, finally, 
\begin{equation} 
\frac{1}{2n\chi}\avg{(x^*-x_0+k+\sqrt{n\Delta \Omega}t)^2}_{t,x_0}= 
\frac{\chi}{2n}\avg{\left(\dudxs\right)^2}_{t,x_0}=\frac{\chi}{2n} 
\left(\frac{\sigma^2}{\Delta}-p^2\right) 
\end{equation} 
\req{hsp} follows immediately. 
 
In order to derive Walras' law, we note that when computing
$\avg{s^*t}_t$, one can make the substitution $t=(\hat\chi
s^*+\epsilon p)/\sigma$ (which is only valid when $s^*>0$). Then
\req{chi} becomes
\begin{equation} 
\chi=\frac{\Delta}{\sigma^2}(n\hat\chi \Omega+\epsilon n\avg{s^*}_t p)= 
\frac{\Delta}{\sigma^2}\left(n\hat\chi \Omega-p^2\chi+kp\right) 
\label{chi1} 
\end{equation} 
where we have used \req{k} in the last equality. Likewise, we can 
substitute $t$ in the average of \req{chihat} by solving \req{xstar} 
for $t$. This yields  
\begin{equation} 
n\hat\chi 
\Omega=\chi\avg{\left(\dudxs\right)^2}_{t,x_0}-kp- \avg{\dudxs 
(x-x_0)}_{t,x_0}  
\end{equation} 
which can be substituted back in \req{chi1}. This yields the desired 
result \req{walras}. 

\section{Almost uniform initial endowments}
\label{appexpdx}

In this section we study the limiting behavior of the economy when the
spread of initial endowments is vanishingly small. In particular we
show that when the initial distribution of endowments becomes uniform
the volume of productive activity vanishes. We take $\Delta=1$ for
simplicity. We take $x_0=\overline{x}_0+\delta x_0$, with
$\overline{x}_0$ a fixed value and $\delta x_0$ a small random
variable, and discuss the solution to the leading order in
$\avg{\delta x^2}$. Then, taking $x^*=\overline{x}_0+\delta x^*$ we
can write, to leading order in $\delta x_0$ and $\delta x^*$
\begin{equation}
\chi \dudxs\cong\chi u'(\overline{x}_0)+\chi u''(\overline{x}_0)\delta x^*\cong
x^*-x_0+k+\sqrt{n\Omega}t=\delta x^*-\delta x_0+\kappa +\sqrt{n\Omega}t
\end{equation} 
From here we can identity the zero and first order terms in $\delta x^*$, viz.
\begin{equation}
\kappa=\chi u'(\overline{x}_0),\quad\delta x^*=
\frac{\delta x_0-\sqrt{n\Omega}t}{1-\chi u''(\overline{x}_0)}
\end{equation}
Then from Eq. \eqref{chihat} we get
\begin{equation}
\widehat{\chi}=\frac{-u''(\overline{x}_0)}{1-\chi u''(\overline{x}_0)}
\end{equation}
and from Eq. \eqref{sigma} 
\begin{equation}
\sigma^2=\left[\frac{u''(\overline{x}_0)}{1-\chi u''(\overline{x}_0)}\right]^2
\left[\avg{(\delta x_0)^2}_{x_0}+n\Omega\right]=\widehat{\chi}^2
\left[\avg{(\delta x_0)^2}_{x_0}+n\Omega\right]
\end{equation}
Coming to the equations for $\Omega$ and $\chi$ we observe that 
$s^*(t)=s_0(t-\tau)\Theta(t-\tau)$ where

\begin{eqnarray}
&s_0=\frac{\sigma}{\widehat{\chi}}=\sqrt{\avg{\delta x_0^2}_{x_0}+n\Omega}\\
&\tau=\frac{\epsilon p}{\sigma}=\frac{\epsilon p}{\widehat{\chi}}
\frac{1}{\sqrt{\avg{\delta x_0^2}_{x_0}+n\Omega}}
\end{eqnarray}
Then we can write $\Omega=s_0^2 I(\tau)$ with
$I(\tau)=\avg{(t-\tau)^2\Theta(t-\tau)}_t$, which can be solved for
$\Omega$
\begin{equation}
\Omega=\frac{\avg{\delta x_0^2 }_{x_0}I(\tau) }{1-nI(\tau)}
\end{equation}
Careful asymptotic analysis implies that in the limit of vanishing
fluctuations $\avg{\delta x_0^2}\to 0$ of the initial endowment $\tau$
diverges as
\begin{equation}
\tau=-\frac{\epsilon u'(\overline{x}_0)}{u''(\overline{x}_0)\avg{\delta x_0^2}_{x_0}^{1/2}}
\end{equation}
Using asymptotic expansion for $I(\tau)$ one finds that the leading
order behavior of $\Omega$ is
\begin{eqnarray}
\Omega\cong\frac{1}{\sqrt{2\pi}}\frac{|u''(\overline{x}_0)|}{\epsilon u'(\overline{x}_0)}
\avg{(\delta x_0)^2}_{x_0}^{3/2}e^{-\frac{\epsilon^2 [u'(\overline{x}_0)]^2}
{2[u''(\overline{x}_0)]^2\avg{(\delta x_0)^2}_{x_0}}}\,.
\end{eqnarray}
Analogously for $\chi$ we have
\begin{eqnarray}
\chi\cong\sqrt{\frac{2}{\pi}}\frac{n}{\epsilon u'(\overline{x}_0)}
\avg{(\delta x_0)^2}_{x_0}^{1/2}e^{-\frac{\epsilon^2 [u'(\overline{x}_0)]^2}
{2[u''(\overline{x}_0)]^2\avg{(\delta x_0)^2}_{x_0}}}\,.
\end{eqnarray}
With these we finally get $\hat\chi\cong |u''(\overline{x}_0)|$ and
$\sigma=|u''(\overline{x}_0)|\sqrt{\avg{\delta x_0^2}_{x_0}}$.
Therefore when the fluctuations of the initial endowment vanish there
is no market activity.

\section{Limit $\epsilon\to 0$}
\label{appexpeps}

Setting $\epsilon=0$ the averages over $s^*(t)$ become trivial and
Eq.s (\ref{Omega},\ref{chi}) and \eqref{k} are easily
evaluated. Progress with the other equations is possible, for generic
$u(x)$ and $\rho(x_0)$, close to $n=2$. Then we expect that the
consumption vector $\boldsymbol{x}$ is nearly constant. Then, as in
the previous case, we assume $x^*=\avg{x_0}+\delta x^*$ with $\delta
x^*$ small. Then $p=u'(\avg{x_0})$ as before. Using the expressions
for $\Omega$, $\chi$ and $\kappa$ and expanding Eq. \eqref{xstar} to
linear order as above, we find the expression
\begin{equation}
\delta x^*=\frac{\frac{2\hat\chi}{n\Delta}(x_0-\avg{x_0})-\sqrt{\frac{2}{n\Delta}}\sigma t}{|u''|+\frac{2\hat\chi}{n\Delta}}
\end{equation}
where $u''=u''(\avg{x_0})$.  This allows us to compute
$\avg{u'(x^*)t}\cong u''\avg{t\delta x^*}$ and hence to evaluate
Eq. \eqref{chihat}:
\begin{equation}
\hat\chi\cong \Delta |u''|\left(1-\frac{n}{2}\right),~~~n\to 2^-
\end{equation}
Likewise we can evaluate Eq. \eqref{sigma} and find
$\sigma=|u''|\sqrt{\Delta(1-n/2)}$. Using these in Eq.s
(\ref{sstar},\ref{Omega}) we find the divergence of $\Omega\cong
1/(2-n)$.

This solution breaks down for $n>2$. We need to take the limit
$\epsilon\to 0$ carefully into account. Again we anticipate that the
spread $\delta x^*$ will be small. More precisely we assume that
$\sigma$ and $\hat\chi$ vanish linearly with $\epsilon$ and set
$s_0=\sigma/\hat\chi$ and $\sigma= d\epsilon$ and look for a solution
with finite $s_0$ and $d$. The existence of such a solution justifies
our assumption.  Then $\Omega=s_0^2 I(p/d)$ where
$I(\tau)=\avg{(t-\tau)^2\Theta(t-\tau)}_t$ has already been introduced
above. The equation for $\chi$ yields $\chi=n\Delta s_0
J(p/d)/(d\epsilon)$ where $J(\tau)=\frac{1}{2}{\rm
erfc}(\tau/\sqrt{2})$. Expanding Eq. \eqref{xstar},as in the previous
section and using the expressions for $\Omega$ and $\chi$ just
derived, we find
\begin{equation}
\delta x^*=\epsilon d\frac{(x_0-\avg{x_0})/s_0-\sqrt{n\Delta I(p/d)}
t}{n\Delta |U''| J(p/d)}
\end{equation}
which justifies our {\em a priori} assumption of small
fluctuations. Inserting this and the expressions of $\Omega$ in
Eq. \eqref{chihat} for $\hat\chi$, after some manipulations, one finds

\begin{equation}\label{nJ}
  n J(p/d)\frac{n}{2}{\rm erfc}\left(\frac{p}{\sqrt{2}d}\right)=1
\end{equation}
which gives $d$ as a function of $n$ and $\avg{x_0}$. Notice that
$\phi=J(p/d)=1/n$. Furthermore $J(\tau)\le 1/2$ for $\tau\ge 0$ which
means that this solution describes the region $n\ge 2$. This equation
also simplifies the expression of $\chi\cong\Delta
s_0/(d\epsilon)$. The equation for $\sigma$ finally gives
\begin{equation}
s_0=\sqrt{\frac{\avg{\delta x_0^2}}{\Delta [1-n I(p/d)]}}
\end{equation}
Note that Eq. \eqref{nJ} implies that $d\to\infty$ as $n\to 2^+$. The
leading behavior is $d\cong \sqrt{\frac{2}{\pi}}\frac{p n}{2-n}$. In
the same limit $I(p/d)\to 1/2$ which means that $\avg{s^*}\propto
s_0\sim 1/(n-2)$ also diverges as $n\to 2^+$ matching the divergence
for $n\to 2^-$.


\begin{thebibliography}{0} 
 
\bibitem{Kirman} A. P. Kirman. Whom or what does the representative 
individual represent. {\em J. Econ. Persp.} {\bf 6} 117 (1992). 
 
\bibitem{Wigner} E. P. Wigner. On the distribution of the roots of 
certain symmetric matrices. {\em Ann. Math.} {\bf 67} 325 (1958). 
 
\bibitem{matching} M. Mezard and G. Parisi. On the solution of the 
random link matching problems. {\em J. Phys. (France)} {\bf 48} 1451 (1987). 
 
\bibitem{ksat} M.Mezard, G.Parisi and R. Zecchina. Analytic and 
algorithmic solution of random satisfiability problems. {\em Science} 
{\bf 297} 812 (2002). 

\bibitem{hkp} J. Hertz, A. Krogh and R.G. Palmer. {\em Introduction to
the theory of neural computation} (Addison-Wesley, Reading, MA, 1991)

\bibitem{cmz} D. Challet, M. Marsili and R. Zecchina. Statistical 
mechanics of heterogeneous agents: minority games.  {\em 
Phys. Rev. Lett.} {\bf 84} 1824 (2000). 
 
\bibitem{bmrz} J. Berg, M. Marsili, A. Rustichini and R. Zecchina. 
Statistical mechanics of asset markets with private information.  {\em 
J. Quant. Fin.} {\bf 1} 203 (2001). 

\bibitem{Foley} D. K. Foley. A statistical equilibrium theory of 
markets. {\em J. Econ. Th.} {\bf 62(2)} 321 (1994). 
 
\bibitem{Durlauf} S. N. Durlauf. How can statistical mechanics 
contribute to social science? {\em Proc. Nat. Acad. Sci.}  {\bf 96} 
10582 (1999). 
 
\bibitem{Follmer} H. Follmer. Random economies with many interacting 
agents. {\em J. Math. Econ.} {\bf 1} 51 (1974). 
 
\bibitem{Lancaster} K. J. Lancaster. {\em Mathematical Economics} 
(Dover, New York, 1987). 
 
\bibitem{Lancaster66} K. J. Lancaster. A new approach to consumer 
theory. {\em J. Pol. Econ.} {\bf 74} 132 (1966). 
 
\bibitem{Takayasu} K. Okuyama, M. Takayasu and H. Takayasu. Zipf's law
in income distribution of companies. {\em Physica A} {\bf 269} 125
(1999).

\bibitem{Axtell} R. L. Axtell. Zipf Distribution of U.S. Firm
Sizes, {\em Science}, {\bf 293}, 1818 (2001).

\bibitem{physics_paper} A. De Martino, M. Marsili and I. P\'erez Castillo in preparation.

\bibitem{Dosi} G. Dosi, {\em Technological Paradigms and Technological Trajectories}, Research Policy, {\bf 11}, 147-162 (1988).

\bibitem{SOC} P. Bak and K. Chen, {\em Self-Organized Criticality} Sci. Am. {\bf 264}, 46-53 (1991).

\bibitem{MPV} M. Mezard, G. Parisi and M. Virasoro. {\em Spin Glass 
theory and beyond} (World Scientific, Singapore, 1987). 
 
\bibitem{talagrandH} M. Talagrand. Rigorous results for the Hopfield
model with many patterns. {\em Prob. Th. Relat. Fields} {\bf 110} 177
(1998).
 
\bibitem{talagrandSG} M. Talagrand. The generalized Parisi
formula. {\em Compt. Rend. Math.} {\bf 337} 111 (2003).
\bibitem{Romer} P. Romer, {\em Endogenous technological change}, J. Pol. Econ. {\bf 98}, S72-S102 (1990).

\end{thebibliography}
\end{document}